\documentclass[twocolumn,letterpaper,aps,prd,superscriptaddress,longbibliography,floatfix]{revtex4-2}
\usepackage{graphicx}	
\usepackage{xspace}     
\usepackage{amsmath}   
\usepackage{hyperref}

\newcommand*\patchAmsMathEnvironmentForLineno[1]{
  \expandafter\let\csname old#1\expandafter\endcsname\csname #1\endcsname
  \expandafter\let\csname oldend#1\expandafter\endcsname\csname end#1\endcsname
  \renewenvironment{#1}
  {\linenomath\csname old#1\endcsname}
  {\csname oldend#1\endcsname\endlinenomath}}
  \newcommand*\patchBothAmsMathEnvironmentsForLineno[1]{
  \patchAmsMathEnvironmentForLineno{#1}
  \patchAmsMathEnvironmentForLineno{#1*}}
  \AtBeginDocument{
  \patchBothAmsMathEnvironmentsForLineno{equation}
  \patchBothAmsMathEnvironmentsForLineno{align}
  \patchBothAmsMathEnvironmentsForLineno{flalign}
  \patchBothAmsMathEnvironmentsForLineno{alignat}
  \patchBothAmsMathEnvironmentsForLineno{gather}
  \patchBothAmsMathEnvironmentsForLineno{multline}
}

\usepackage{lineno}   
\usepackage{amsmath}  

\begin{document}

\title{Improving constraints on gluon spin-momentum correlations in transversely
polarized protons via midrapidity open-heavy-flavor electrons in
$p^{\uparrow}+p$ collisions at $\sqrt{s}=200$ GeV} 

\newcommand{\abilene}{Abilene Christian University, Abilene, Texas 79699, USA}
\newcommand{\augie}{Department of Physics, Augustana University, Sioux Falls, South Dakota 57197, USA}
\newcommand{\banaras}{Department of Physics, Banaras Hindu University, Varanasi 221005, India}
\newcommand{\barc}{Bhabha Atomic Research Centre, Bombay 400 085, India}
\newcommand{\baruch}{Baruch College, City University of New York, New York, New York, 10010 USA}
\newcommand{\bnlcoll}{Collider-Accelerator Department, Brookhaven National Laboratory, Upton, New York 11973-5000, USA}
\newcommand{\bnlphys}{Physics Department, Brookhaven National Laboratory, Upton, New York 11973-5000, USA}
\newcommand{\caucr}{University of California-Riverside, Riverside, California 92521, USA}
\newcommand{\charlesczech}{Charles University, Ovocn\'{y} trh 5, Praha 1, 116 36, Prague, Czech Republic}
\newcommand{\cns}{Center for Nuclear Study, Graduate School of Science, University of Tokyo, 7-3-1 Hongo, Bunkyo, Tokyo 113-0033, Japan}
\newcommand{\colorado}{University of Colorado, Boulder, Colorado 80309, USA}
\newcommand{\columbia}{Columbia University, New York, New York 10027 and Nevis Laboratories, Irvington, New York 10533, USA}
\newcommand{\czechtech}{Czech Technical University, Zikova 4, 166 36 Prague 6, Czech Republic}
\newcommand{\debrecen}{Debrecen University, H-4010 Debrecen, Egyetem t{\'e}r 1, Hungary}
\newcommand{\elte}{ELTE, E{\"o}tv{\"o}s Lor{\'a}nd University, H-1117 Budapest, P{\'a}zm{\'a}ny P.~s.~1/A, Hungary}
\newcommand{\ewha}{Ewha Womans University, Seoul 120-750, Korea}
\newcommand{\famu}{Florida A\&M University, Tallahassee, FL 32307, USA}
\newcommand{\fsu}{Florida State University, Tallahassee, Florida 32306, USA}
\newcommand{\gsu}{Georgia State University, Atlanta, Georgia 30303, USA}
\newcommand{\hiroshima}{Physics Program and International Institute for Sustainability with Knotted Chiral Meta Matter (SKCM2), Hiroshima University, Higashi-Hiroshima, Hiroshima 739-8526, Japan}
\newcommand{\howard}{Department of Physics and Astronomy, Howard University, Washington, DC 20059, USA}
\newcommand{\ihepprot}{IHEP Protvino, State Research Center of Russian Federation, Institute for High Energy Physics, Protvino, 142281, Russia}
\newcommand{\illuiuc}{University of Illinois at Urbana-Champaign, Urbana, Illinois 61801, USA}
\newcommand{\inrras}{Institute for Nuclear Research of the Russian Academy of Sciences, prospekt 60-letiya Oktyabrya 7a, Moscow 117312, Russia}
\newcommand{\instpasczech}{Institute of Physics, Academy of Sciences of the Czech Republic, Na Slovance 2, 182 21 Prague 8, Czech Republic}
\newcommand{\isu}{Iowa State University, Ames, Iowa 50011, USA}
\newcommand{\jaea}{Advanced Science Research Center, Japan Atomic Energy Agency, 2-4 Shirakata Shirane, Tokai-mura, Naka-gun, Ibaraki-ken 319-1195, Japan}
\newcommand{\jeonbuk}{Jeonbuk National University, Jeonju, 54896, Korea}
\newcommand{\kek}{KEK, High Energy Accelerator Research Organization, Tsukuba, Ibaraki 305-0801, Japan}
\newcommand{\korea}{Korea University, Seoul 02841, Korea}
\newcommand{\kurchatov}{National Research Center ``Kurchatov Institute", Moscow, 123098 Russia}
\newcommand{\kyoto}{Kyoto University, Kyoto 606-8502, Japan}
\newcommand{\lawllnl}{Lawrence Livermore National Laboratory, Livermore, California 94550, USA}
\newcommand{\losalamos}{Los Alamos National Laboratory, Los Alamos, New Mexico 87545, USA}
\newcommand{\lund}{Department of Physics, Lund University, Box 118, SE-221 00 Lund, Sweden}
\newcommand{\lyon}{IPNL, CNRS/IN2P3, Univ Lyon, Université Lyon 1, F-69622, Villeurbanne, France}
\newcommand{\maryland}{University of Maryland, College Park, Maryland 20742, USA}
\newcommand{\mass}{Department of Physics, University of Massachusetts, Amherst, Massachusetts 01003-9337, USA}
\newcommand{\mate}{MATE, Laboratory of Femtoscopy, K\'aroly R\'obert Campus, H-3200 Gy\"ongy\"os, M\'atrai\'ut 36, Hungary}
\newcommand{\michigan}{Department of Physics, University of Michigan, Ann Arbor, Michigan 48109-1040, USA}
\newcommand{\miss}{Mississippi State University, Mississippi State, Mississippi 39762, USA}
\newcommand{\muhlenberg}{Muhlenberg College, Allentown, Pennsylvania 18104-5586, USA}
\newcommand{\nara}{Nara Women's University, Kita-uoya Nishi-machi Nara 630-8506, Japan}
\newcommand{\natmephi}{National Research Nuclear University, MEPhI, Moscow Engineering Physics Institute, Moscow, 115409, Russia}
\newcommand{\newmex}{University of New Mexico, Albuquerque, New Mexico 87131, USA}
\newcommand{\nmsu}{New Mexico State University, Las Cruces, New Mexico 88003, USA}
\newcommand{\northcg}{Physics and Astronomy Department, University of North Carolina at Greensboro, Greensboro, North Carolina 27412, USA}
\newcommand{\ohio}{Department of Physics and Astronomy, Ohio University, Athens, Ohio 45701, USA}
\newcommand{\ornl}{Oak Ridge National Laboratory, Oak Ridge, Tennessee 37831, USA}
\newcommand{\orsay}{IPN-Orsay, Univ.~Paris-Sud, CNRS/IN2P3, Universit\'e Paris-Saclay, BP1, F-91406, Orsay, France}
\newcommand{\peking}{Peking University, Beijing 100871, People's Republic of China}
\newcommand{\pnpi}{PNPI, Petersburg Nuclear Physics Institute, Gatchina, Leningrad region, 188300, Russia}
\newcommand{\pusan}{Pusan National University, Pusan 46241, Korea}
\newcommand{\riken}{RIKEN Nishina Center for Accelerator-Based Science, Wako, Saitama 351-0198, Japan}
\newcommand{\rikjrbrc}{RIKEN BNL Research Center, Brookhaven National Laboratory, Upton, New York 11973-5000, USA}
\newcommand{\rikkyo}{Physics Department, Rikkyo University, 3-34-1 Nishi-Ikebukuro, Toshima, Tokyo 171-8501, Japan}
\newcommand{\saispbstu}{Saint Petersburg State Polytechnic University, St.~Petersburg, 195251 Russia}
\newcommand{\seoulnat}{Department of Physics and Astronomy, Seoul National University, Seoul 151-742, Korea}
\newcommand{\stonybrkc}{Chemistry Department, Stony Brook University, SUNY, Stony Brook, New York 11794-3400, USA}
\newcommand{\stonycrkp}{Department of Physics and Astronomy, Stony Brook University, SUNY, Stony Brook, New York 11794-3800, USA}
\newcommand{\tenn}{University of Tennessee, Knoxville, Tennessee 37996, USA}
\newcommand{\texsu}{Texas Southern University, Houston, TX 77004, USA}
\newcommand{\titech}{Department of Physics, Tokyo Institute of Technology, Oh-okayama, Meguro, Tokyo 152-8551, Japan}
\newcommand{\tsukuba}{Tomonaga Center for the History of the Universe, University of Tsukuba, Tsukuba, Ibaraki 305, Japan}
\newcommand{\vandy}{Vanderbilt University, Nashville, Tennessee 37235, USA}
\newcommand{\weizmann}{Weizmann Institute, Rehovot 76100, Israel}
\newcommand{\wigner}{Institute for Particle and Nuclear Physics, Wigner Research Centre for Physics, Hungarian Academy of Sciences (Wigner RCP, RMKI) H-1525 Budapest 114, POBox 49, Budapest, Hungary}
\newcommand{\yonsei}{Yonsei University, IPAP, Seoul 120-749, Korea}
\newcommand{\zagreb}{Department of Physics, Faculty of Science, University of Zagreb, Bijeni\v{c}ka c.~32 HR-10002 Zagreb, Croatia}
\newcommand{\zambia}{Department of Physics, School of Natural Sciences, University of Zambia, Great East Road Campus, Box 32379, Lusaka, Zambia}
\affiliation{\abilene}
\affiliation{\augie}
\affiliation{\banaras}
\affiliation{\barc}
\affiliation{\baruch}
\affiliation{\bnlcoll}
\affiliation{\bnlphys}
\affiliation{\caucr}
\affiliation{\charlesczech}
\affiliation{\cns}
\affiliation{\colorado}
\affiliation{\columbia}
\affiliation{\czechtech}
\affiliation{\debrecen}
\affiliation{\elte}
\affiliation{\ewha}
\affiliation{\famu}
\affiliation{\fsu}
\affiliation{\gsu}
\affiliation{\hiroshima}
\affiliation{\howard}
\affiliation{\ihepprot}
\affiliation{\illuiuc}
\affiliation{\inrras}
\affiliation{\instpasczech}
\affiliation{\isu}
\affiliation{\jaea}
\affiliation{\jeonbuk}
\affiliation{\kek}
\affiliation{\korea}
\affiliation{\kurchatov}
\affiliation{\kyoto}
\affiliation{\lawllnl}
\affiliation{\losalamos}
\affiliation{\lund}
\affiliation{\lyon}
\affiliation{\maryland}
\affiliation{\mass}
\affiliation{\mate}
\affiliation{\michigan}
\affiliation{\miss}
\affiliation{\muhlenberg}
\affiliation{\nara}
\affiliation{\natmephi}
\affiliation{\newmex}
\affiliation{\nmsu}
\affiliation{\northcg}
\affiliation{\ohio}
\affiliation{\ornl}
\affiliation{\orsay}
\affiliation{\peking}
\affiliation{\pnpi}
\affiliation{\pusan}
\affiliation{\riken}
\affiliation{\rikjrbrc}
\affiliation{\rikkyo}
\affiliation{\saispbstu}
\affiliation{\seoulnat}
\affiliation{\stonybrkc}
\affiliation{\stonycrkp}
\affiliation{\tenn}
\affiliation{\texsu}
\affiliation{\titech}
\affiliation{\tsukuba}
\affiliation{\vandy}
\affiliation{\weizmann}
\affiliation{\wigner}
\affiliation{\yonsei}
\affiliation{\zagreb}
\affiliation{\zambia}
\author{N.J.~Abdulameer} \affiliation{\debrecen}
\author{U.~Acharya} \affiliation{\gsu} 
\author{C.~Aidala} \affiliation{\michigan} 
\author{Y.~Akiba} \email[PHENIX Spokesperson: ]{akiba@rcf.rhic.bnl.gov} \affiliation{\riken} \affiliation{\rikjrbrc} 
\author{M.~Alfred} \affiliation{\howard} 
\author{V.~Andrieux} \affiliation{\michigan} 
\author{N.~Apadula} \affiliation{\isu} 
\author{H.~Asano} \affiliation{\kyoto} \affiliation{\riken} 
\author{B.~Azmoun} \affiliation{\bnlphys} 
\author{V.~Babintsev} \affiliation{\ihepprot} 
% ...
\collaboration{PHENIX Collaboration}  \noaffiliation

\date{\today}


\begin{abstract}


Polarized proton-proton collisions provide leading-order access to 
gluons, presenting an opportunity to constrain gluon spin-momentum 
correlations within transversely polarized protons and enhance our 
understanding of the three-dimensional structure of the proton.  
Midrapidity open-heavy-flavor production at $\sqrt{s}=200$ GeV is 
dominated by gluon-gluon fusion, providing heightened sensitivity to 
gluon dynamics relative to other production channels. Transverse 
single-spin asymmetries of positrons and electrons from heavy-flavor 
hadron decays are measured at midrapidity using the PHENIX detector at 
the Relativistic Heavy Ion Collider. These charge-separated measurements 
are sensitive to gluon correlators that can in principle be related to 
gluon orbital angular momentum via model calculations.  Explicit 
constraints on gluon correlators are extracted for two separate models, 
one of which had not been constrained previously.

\end{abstract}

\maketitle



\section{Introduction}

Polarized proton-proton collisions provide a unique opportunity to 
improve our understanding of gluon contributions to the spin structure 
of the proton, because they are accessible at leading order, which 
is not true for lepton-hadron scattering. The complex spin structure of the 
proton leads to emergent properties such as spin-momentum and spin-spin 
correlations analogous to the fine and hyperfine structure of atoms. 
These correlations in protons are experimentally accessible through 
observables known as transverse single-spin asymmetries (TSSAs).  TSSAs 
quantify azimuthal modulations of particle production in collisions of 
transversely polarized nucleons with unpolarized particles, and have 
been measured to reach magnitudes up to 40\% in hadron-hadron 
collisions~\cite{Klem:1976ui, Adams:1991cs, Allgower:2002qi, 
Arsene:2008aa}. Perturbative quantum chromodynamics (pQCD) calculations 
had predicted TSSAs of $<1$\% from purely perturbative 
contributions~\cite{pQCDTSSA}; recent calculations suggest small 
additional perturbative contributions~\cite{twoLoopsTSSA}.

Two complementary theoretical frameworks exist for describing large 
TSSAs in which contributions arise from \emph{nonperturbative} elements 
of the factorized cross section --- transverse-momentum-dependent (TMD) 
factorization~\cite{SiversTMD,BoerMuldersTMD, CollinsTMDFF}, and twist-3 
factorization~\cite{Efremov:1984ip, SSA_twist3} (see 
Ref.~\cite{TSSA_review} for a recent review). The two frameworks are 
related, and phenomenological arguments indicate TSSAs in various 
reactions share a common origin in multiparton 
correlations~\cite{SSAorigin}. The TMD framework has explicit dependence 
on transverse momentum $k_{T}$ of partons within hadrons in addition to 
the longitudinal momentum fraction $x$. In this approach, standard 
collinear parton distribution functions (PDFs) and fragmentation 
functions (FFs) are replaced with TMD functions. The twist-3 approach 
considers power-suppressed terms with respect to the hard-scattering 
energy scale $Q$ in the factorization expansion. Constraining TMD 
functions experimentally requires access to both a hard scale $Q$ and 
soft scale $k_T$ sensitive to partonic transverse momentum in the proton 
or the process of hadronization, with $Q \gg k_T$, while the 
higher-twist formalism only requires access to a hard scale that is 
represented by the transverse momentum of the produced particle 
($p_{T}$). Twist-3 correlation functions can be written in terms of 
$k_{T}$ moments of corresponding TMDs~\cite{Ji:2006ub}. Both frameworks 
have demonstrated success in modeling TSSAs in complementary regions of 
$p_{T}$~\cite{Ji:2006ub,TMD_twist3_compare, Collins_twist3_compare}, and 
are relevant for constraining orbital angular momentum of quarks and 
gluons in protons~\cite{orbital1, orbital2, orbital3}. At twist-3, 
quantum interference between standard $2 \rightarrow 2$ QCD processes 
and some processes involving an extra gluon must be considered, 
introducing additional terms to cross-section calculations depending on 
the number of colliding or produced hadrons. These terms encode quantum 
interference in twist-3 correlation functions convoluted with standard 
collinear PDFs and FFs. TSSAs are defined in Eq.~(\ref{eqn:AN_phi}), 
leading to the following proportionality at 
twist-3~\cite{TSSA_twist3_eqn_intro, TSSA_eqn}:
\begin{align}
\begin{small}
	\begin{aligned}
	    A_{N} \propto	\sum_{a,b,c}\phi^{(3)}_{a/A} (x_{1},x_{2},\vec{s}_{\perp}) \otimes \phi_{b/B} (x') \otimes \hat{\sigma} \otimes D_{c \rightarrow C}(z) \\
	    + \sum_{a,b,c} \delta q_{a/A} (x,\vec{s}_{\perp}) \otimes\phi^{(3)}_{b/B} (x'_{1},x'_{2}) \otimes \hat{\sigma} '\otimes D_{c \rightarrow C} (z) \\ 
	    + \sum_{a,b,c} \delta q_{a/A} (x,\vec{s}_{\perp}) \otimes \phi_{b/B} (x') \otimes \hat{\sigma} '' \otimes D^{(3)}_{c \rightarrow C} (z_{1},z_{2})	   
	\end{aligned}      
\label{eqn:AN_twist3}  	
\end{small}
\end{align}
\noindent Each term with a superscript $(3)$ corresponds to a twist-3 
correlation function; the rest are at leading twist (twist-2), where 
$\otimes$ represents a convolution in longitudinal momentum fractions 
($x$) of partons in parent protons and collinear momentum fractions 
($z$) of produced hadrons with respect to their originating 
partons~\cite{TSSA_eqn}. The primed variables originate from the 
unpolarized proton in the initial state, and the numbered variables 
appear in twist-3 correlators, where multiparton correlations must be 
considered. The $\phi$ and $D$ denote PDFs and FFs respectively, where the 
lowercase subscripts represent the parton type, and the uppercase 
subscripts represent the parent hadron. The term $\delta 
q_{x/X}(x,\vec{s}_{\perp})$ is the transversity distribution, a 
spin-spin correlation of transversely polarized quarks in transversely 
polarized hadrons~\cite{transversityDef}. Twist-3 correlators have more 
intuitive physical meaning through their relation to corresponding 
TMDs~\cite{Ji:2006ub, SiversTMD, BoerMuldersTMD, CollinsTMDFF}.

In $p$$+$$p$ collisions at $\sqrt{s}=200$~GeV, open-heavy-flavor (OHF) 
production at midrapidity is dominated by gluon-gluon fusion, receiving 
only a small contribution from quark-antiquark 
annihilation~\cite{heavy_quark_production}. In gluon-gluon fusion 
events, only the first term in Eq.~(\ref{eqn:AN_twist3}) is relevant (as 
the gluon does not have a transversity distribution in spin $1/2$ 
nucleons), providing sensitivity to the trigluon correlation functions 
in polarized protons. The relevant twist-3 correlators for 
quark-antiquark annihilation and gluon-gluon fusion are the 
Efremov-Teryaev-Qiu-Sterman ($qgq$) 
correlator~\cite{Efremov:1984ip,Qiu:1991wg}, and the trigluon ($ggg$) 
correlators~\cite{trigluon_intro, trigluon_clar1, trigluon_clar2, 
trigluon_twists_Yoshida_SIDIS, trigluon_twists_SIDIS_kang} respectively. 
Note that the trigluon correlators were introduced in 
Ref.~\cite{trigluon_intro}, and were subsequently clarified to be two 
independent 
functions~\cite{trigluon_clar1,trigluon_clar2,trigluon_twists_Yoshida_SIDIS,trigluon_twists_SIDIS_kang}.  
The $qgq$ correlator has been experimentally constrained from global 
fits, discussed in Ref.~\cite{SSAorigin}, while the $ggg$ correlators 
have received less attention, with few measurements capable of providing 
indirect 
constraints~\cite{STAR_inclusivejet_TSSA1,AnDY_gluons,SIDIS_gluonSivers,STAR_inclusivejet_TSSA2,PPG234,STAR_inclusive_jets}
or direct constraints~\cite{ohf_muons,PPG235}.

The TSSA for open-charm production in $p^{\uparrow}+p$ collisions at 
$\sqrt{s}=200$ GeV was calculated in Refs.~\cite{trigluon_twists} 
and~\cite{trigluon_twists_Yoshida} within the twist-3 framework. The 
trigluon correlation functions are defined in 
Ref.~\cite{trigluon_twists} as $T_{G}^{(f)}(x,x)$ (antisymmetric) and 
$T_{G}^{(d)}(x,x)$ (symmetric), where the $(f)$ and $(d)$ superscripts 
represent three gluon-field color indices contracting with antisymmetric 
or symmetric structure tensors. Lack of direct information on the 
trigluon correlators has led to simple phenomenological models with 
normalization parameters to the unpolarized gluon PDF. In 
Ref.~\cite{trigluon_twists} (following from 
Ref.~\cite{TSSA_twist3_eqn_intro}) parameters $\lambda_{f}$ and 
$\lambda_{d}$ are introduced:
\begin{equation}
T_{G}^{(f)}(x,x) = \lambda_{f} G(x), \hspace{3em}    T_{G}^{(d)}(x,x) 
= \lambda_{d} G(x). \label{eqn:gluonNorms}
\end{equation} 
The trigluon correlation functions in 
Ref.~\cite{trigluon_twists_Yoshida} are instead defined as $N(x_{1}, 
x_{2})$ (antisymmetric), and $O(x_{1}, x_{2})$ (symmetric), with four 
independent contributions to TSSAs, $\{N(x,x), N(x,0), O(x,x),O(x,0)\}$. 
As shown in Ref.~\cite{trigluon_twists_Yoshida}, at $\sqrt{s}=200$ GeV 
the asymmetries depend on effective trigluon correlators $N(x,x) - 
N(x,0)$ and $O(x,x) + O(x,0)$, which are directly related to 
$T_{G}^{(f)}$ and $T_{G}^{(d)}$ in 
Ref.~\cite{trigluon_twists_Yoshida_SIDIS}. 
Reference~\cite{trigluon_twists_Yoshida} introduces parameters $K_{G}$ 
and $K_{G}'$ with the assumptions:

\begin{align}
O(x,x) &= O(x,0) = N(x,x) = -N(x,0)\label{eqn:conditions} \\
[{\rm Model 1}]\ \ O(x,x) &= K_{G} x G(x)\label{eqn:model1} \\
[{\rm Model 2}]\ \ O(x,x) &= K_{G}' \sqrt{x} G(x)\label{eqn:model2} 
\end{align}

Note that the assumptions on the trigluon correlators in 
Eqs.~\ref{eqn:gluonNorms},~\ref{eqn:model1}, and~\ref{eqn:model2}
(e.g., the functional dependence on $x$ and the proportionality to the 
unpolarized gluon PDF) are oversimplified.  For this reason, it is 
advantageous to compare to different models with various $x$ dependencies. 
The results presented in this paper place direct constraints on 
$\lambda_{f}, \lambda_{d}, K_{G}$ and $K_{G}'$.

Open-charm production at the Relativistic Heavy Ion Collider (RHIC) has 
also been investigated with the TMD factorization approach as a means of 
constraining the gluon Sivers PDF (see 
Refs.~\cite{gluonSivers,gluonSivers2,gluonSivers3}). The measurements 
presented here will be useful in providing constraints to the gluon 
Sivers TMD PDF through constraining the twist-3 trigluon correlators, 
which are related to $k_{T}$ moments of the gluon Sivers 
PDF~\cite{Ji:2006ub}.

\section{Data Analysis}

The asymmetry measurements presented here utilize data recorded in 2015 
by the PHENIX experiment at RHIC with collisions of transversely 
polarized protons on transversely polarized protons at $\sqrt{s} = 200$ 
GeV, and approximately 23 pb$^{-1}$ of integrated luminosity. The 
polarization of each beam at RHIC in 2015 is measured to be $0.58 \pm 
0.02$ for the clockwise beam and $0.60 \pm 0.02$ for the 
counterclockwise beam, with transverse polarization direction aligned 
vertically to the accelerator plane~\cite{RHIC_polarimetry}. The 
polarization direction is varied from bunch to bunch (a) to reduce 
systematics related to detector coverage and performance, and (b) to 
allow for the polarization of a single beam to be considered at a time 
by averaging over the polarization directions of the opposing beam.  This 
yields two independent data sets from which the transverse single-spin 
asymmetries are extracted, validated for consistency, and averaged to 
obtain the final result.

The PHENIX detector is described in detail in Ref.~\cite{PHENIX_det}. 
Detector subsystems used for midrapidity charged-particle detection 
comprise two central-arm spectrometers oriented to the left and right of 
the beam axis, each with acceptance $|\eta| < 0.35$ and $\Delta \phi = 
0.5\pi$, and a silicon vertex detector (VTX)~\cite{VTX_ref, VTX_status} 
with acceptance of $|\eta| < 1$ and $\Delta \phi \approx 0.8 \pi$ per 
arm. The central arms contain drift and pad chambers for 
tracking~\cite{tracking_ref}, electromagnetic calorimeters (EMCal) to 
measure energy deposition of charged particles and 
photons~\cite{EMCal_ref}, and a ring-imaging \v{C}erenkov (RICH) 
detector for particle identification with $e/\pi$ separation up to $5$ 
GeV/$c$~\cite{RICH_ref}.

Curating the electron candidate sample follows the same procedure as in 
Ref.~\cite{PPG223}. The electron candidate sample is composed of tracks 
reconstructed from hits in the drift and pad chambers of the central arm 
spectrometers coincident with hits in the silicon vertex detector.  
Tracks within $1.0 < p_{T}$ (GeV/c) $< 5.0$ that fire at least one 
photomultiplier (PMT) tube in the RICH detector, and that have a maximum 
displacement of 5 cm between the track projection and center of the ring 
of \v{C}erenkov light as measured by the PMTs in the RICH are 
considered. In order to increase the electron purity, track energy $E$ 
deposited in the EMCal and track momentum $p$ should have a ratio near 
unity, as electrons deposit most of their energy in the EMCal while 
charged hadrons do not. The $E/p$ distribution for electron candidates 
in Run-15 was fit with an exponential + Gaussian, where the mean 
$\mu_{E/p}$ and width $\sigma_{E/p}$ of the Gaussian portion were 
extracted and used to impose the following condition $\left| (E/p - 
\mu_{E/p})/\sigma_{E/p}\right| < 2$. Spatial displacements $\Delta z$ 
and $\Delta \phi$ of track projections and corresponding electromagnetic 
showers in the EMCal are required to be separated by no more than 3 
standard deviations of the corresponding $\Delta z$ and $\Delta \phi$ 
distributions, and the probability that an EMCal cluster originates from 
an electromagnetic shower (as calculated by the shower shape) is 
required to be above 0.01. Tracks reconstructed in the central arms are 
projected to the VTX detector and fit to coincidental VTX hits via the 
iterative algorithm described in Ref.~\cite{PHENIX:2015ynp} --- the fit 
is required to satisfy $\chi^{2}/ndf < 3$. A hit is required in both of 
the inner two layers of the VTX to veto conversion electrons created by 
photons interacting with detector material, and an additional hit is 
required in either of the outer layers of the VTX. The narrow opening 
angle between $e^{+}e^{-}$ from photonic conversions is exploited to 
further reduce background from conversions in the beam pipe or inner two 
layers of the VTX; more details for this and the VTX detector can be 
found in Ref.~\cite{PPG223}. An additional requirement was placed on the 
number of live trigger counts per bunch crossing because the asymmetry 
analysis is performed bunch-by-bunch.

TSSAs can be calculated as amplitudes of sinusoidal modulations 
of azimuthal particle production:
\begin{equation}
A_{N} (\phi) = \frac{\sigma^{\uparrow}(\phi) - \sigma^{\downarrow} (\phi)}{\sigma^{\uparrow}(\phi) + \sigma^{\downarrow} (\phi)} = A_{N} \cos\phi,
\label{eqn:AN_phi}
\end{equation}

\noindent where $\sigma^{\uparrow, \downarrow}(\phi)$ correspond to 
transversely polarized cross sections for different spin orientations. 
Due to the nature of the azimuthal angular acceptance of the PHENIX 
spectrometer arms, the measurements of midrapidity TSSAs are integrated 
in $\phi$ for one arm at a time. This necessitates division by an 
azimuthal correction factor $\langle\mid\cos\phi\mid\rangle$. 
Equation~(\ref{eqn:AN_phi}) must also be corrected for the polarization 
$P$. All of these corrections are applied as seen in the ``relative 
luminosity formula'', a well-established PHENIX method used in 
Refs.~\cite{PPG235, PPG234, ohf_muons, PPG103E, PPG050} to extract 
TSSAs:
\begin{equation}
A_{N} = \frac{1}{P\langle\mid\cos\phi\mid\rangle }
\frac{N^{\uparrow}-\mathcal{R} N^{\downarrow}}
{N^{\uparrow}+\mathcal{R} N^{\downarrow}}
\label{eqn:AN_lumi}.
\end{equation}

In Eq.~(\ref{eqn:AN_lumi}), $N^{\uparrow, \downarrow}$ are the 
spin-dependent yields for collisions with $\uparrow, \downarrow$ 
polarized bunch crossings respectively, and $\mathcal{R} = 
\mathcal{L}^{\uparrow}/\mathcal{L}^{\downarrow}$ is the relative 
luminosity, defined as the ratio of luminosities for collisions with 
oppositely oriented bunch crossing polarization. The azimuthal 
correction factor $\langle\mid\cos\phi\mid\rangle$ is calculated in 
each transverse momentum ($p_{T}$) bin for the electron candidate sample 
to account for detector efficiency effects. To serve as a cross check to 
Eq.~(\ref{eqn:AN_lumi}), the asymmetries are also calculated with the 
``square root formula,'' as shown in Eq.~(\ref{eqn:AN_sqrt}). The 
difference in asymmetries calculated with the separate methods is taken 
as a systematic uncertainty $\sigma^{\rm sys}_{\rm diff}$.

\begin{equation}
A_{N} = \frac{1}{P\langle\mid\cos\phi\mid\rangle }\frac{\sqrt{N^{\uparrow}_{L}N^{\downarrow}_{R}} - \sqrt{N^{\downarrow}_{L}N^{\uparrow}_{R}}}{\sqrt{N^{\uparrow}_{L}N^{\downarrow}_{R}} + \sqrt{N^{\downarrow}_{L}N^{\uparrow}_{R}}}\label{eqn:AN_sqrt}.
\end{equation}
The $L,R$ subscripts represent the left and right spectrometer arm with 
respect to the polarized proton-going direction. The square root formula 
cannot be used independently on each spectrometer arm, leading to only 
two independent data sets for cross validation and averaging 
corresponding to the two beams, rather than four independent data sets 
as is the case for the relative luminosity formula, corresponding to the 
two beams and two spectrometer arms. As an additional cross check, 
$A_{N}$ was calculated as shown in Eq.~\ref{eqn:AN_phi}, via sinusoidal 
fits, with 3 $\phi$ bins per spectrometer arm, yielding consistent 
results with that of Eqs.~\ref{eqn:AN_lumi} and~\ref{eqn:AN_sqrt}.

\begin{figure}[htb]		
        \includegraphics[width=1.0\linewidth]{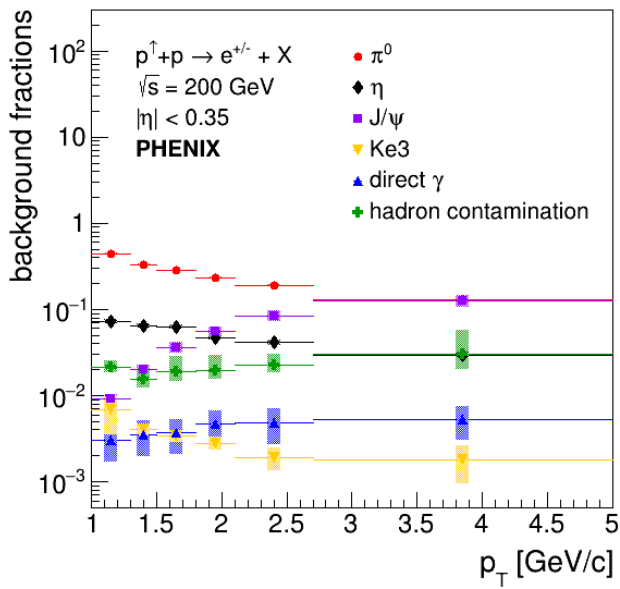}
        \caption{Fraction of measured electron candidates attributed to 
each background source ($f_{i}$) -- charge combined (+/-)}
        \label{fig:BGFractionPlot}
    \end{figure}

\begin{figure*}[htb]		
        \begin{minipage}{0.48\linewidth}
\includegraphics[width=0.99\linewidth]{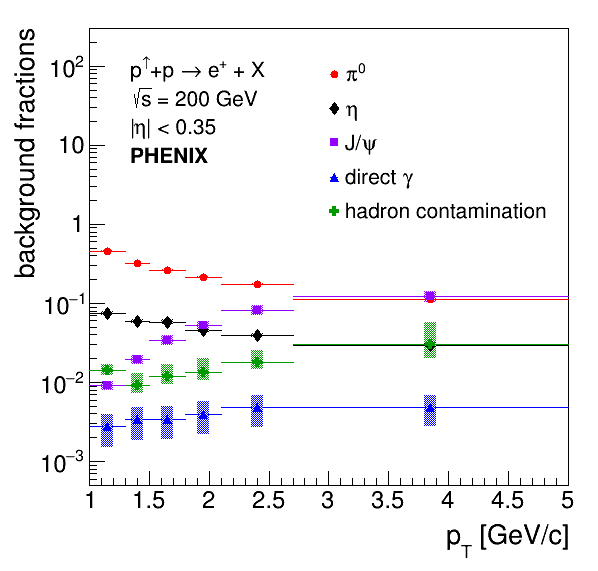}
            \caption{Fractions of measured positron candidates 
attributed to each background source ($f_{i}$); results are shown in 
Table~\ref{tab:bgFracs} and used as inputs to the background correction 
procedure -- charge (+)}
            \label{fig:BGFractionPlotp}
        \end{minipage}
        \begin{minipage}{0.48\linewidth}
\includegraphics[width=0.99\linewidth]{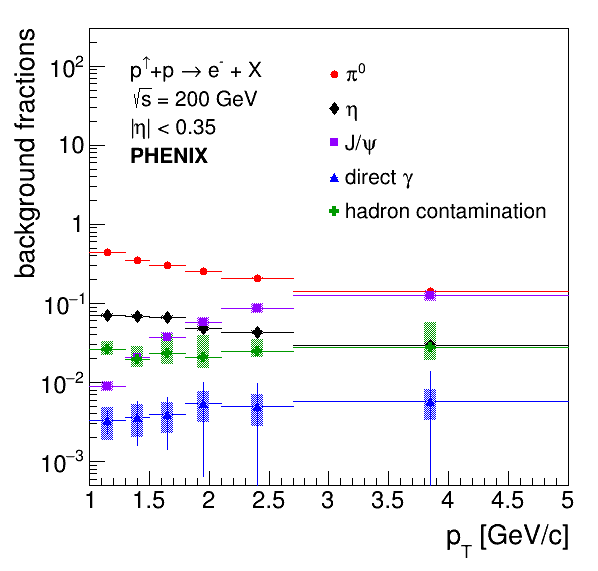}
            \caption{Fractions of measured electron candidates 
attributed to each background source ($f_{i})$; results are shown in 
Table~\ref{tab:bgFracs} and used as inputs to the background correction 
procedure -- charge (-)}
            \label{fig:BGFractionPlotm}
        \end{minipage}
\begin{minipage}{0.48\linewidth}
    \includegraphics[width=0.99\linewidth]{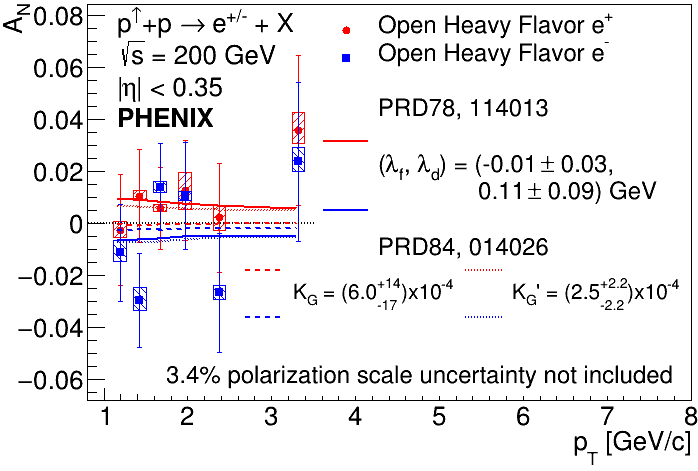}
    \caption{$A_{N}(OHF\rightarrow{e^{\pm}})$ (red) circles and (blue) 
squares for positrons and electrons, respectively. Also plotted are 
predictions of $A_{N}(D^{0}/\bar{D^{0}}\rightarrow{e^{\pm}})$ from 
Ref.~\cite{trigluon_twists}, and 
$A_{N}((D^{0}/\bar{D^{0}}+D^{+/-})\rightarrow e^{\pm})$ from 
Ref.~\cite{trigluon_twists_Yoshida} for best-fit 
trigluon-correlator-normalization parameters, with the red/blue solid, 
dashed, and dotted lines corresponding to central values of the $1 
\sigma$ confidence intervals shown in the legend.}
    \label{fig:finalANs}
\end{minipage}
\begin{minipage}{0.48\linewidth}
    \includegraphics[width=0.99\linewidth]{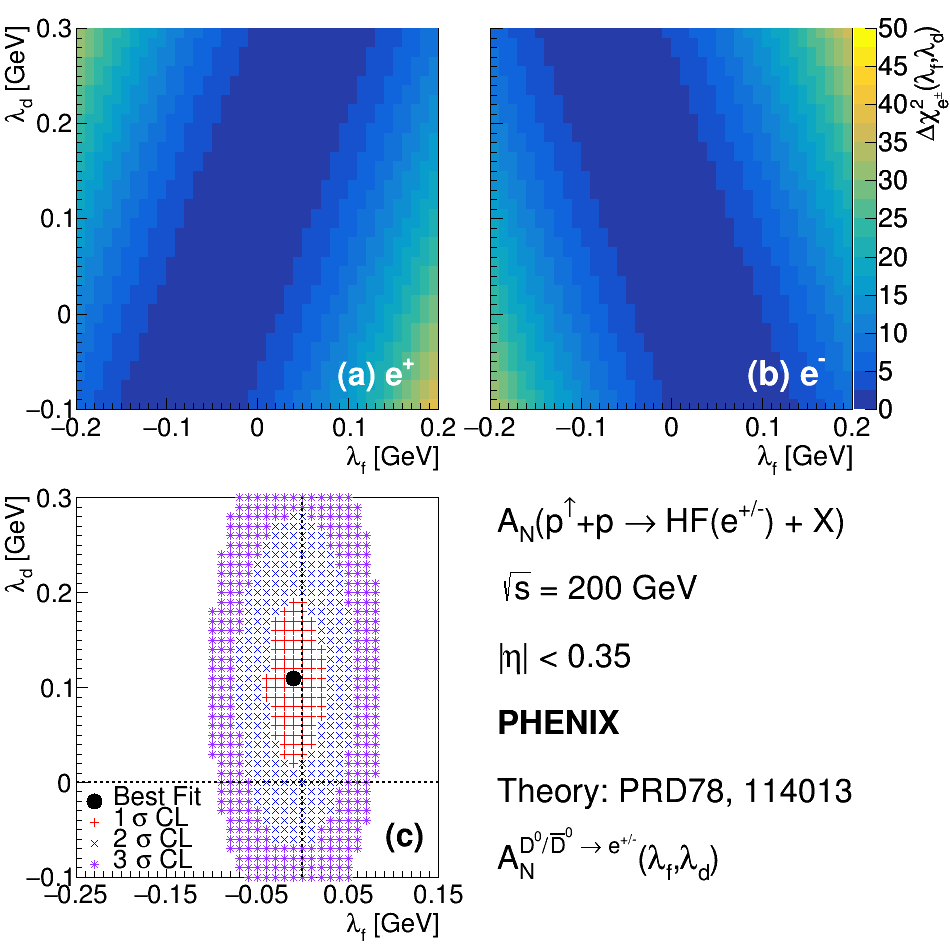}
    \caption{Results of the statistical analysis performed to extract 
best-fit parameters $\lambda_{f}$ and $\lambda_{d}$ by comparing data to 
theory~\cite{trigluon_twists}. $\chi^{2}(\lambda_{f}, \lambda_{d}) - 
\chi^{2}_{\rm min}$ is shown for (a) $e^{+}$ and (b) $e^{-}$. 
Panel (c) shows the 1, 2, and 3$\sigma$ confidence level regions, 
$\chi^{2}(\lambda_{f}, \lambda_{d}) - \chi^{2}_{\rm min} < n^{2}$ 
($n = 1, 2, 3$).}
    \label{fig:LamParam}
\end{minipage}
\end{figure*}

Once $A_{N}$ is calculated for the electron candidate sample, 
background corrections allow for extraction of the asymmetry for OHF decay 
electrons.  The relevant background sources are electrons from other 
parent particles ($\pi^{0}, \eta$, direct photons $\gamma, J/\psi, 
K^{0}_{S}, K^{\pm}$) and charged hadrons misidentified as electrons 
(primarily $\pi^{\pm}$). To calculate the background-corrected 
asymmetry, the fraction of each background source present in the data 
sample needs to be calculated and the background asymmetries need to be 
measured. Equation~(\ref{eqn:BackgroundCorrection}) shows the formula 
for extracting the ${\rm OHF}\rightarrow{e}$ asymmetry from the 
electron-candidate-sample asymmetry,
\begin{equation}
    A_{N}^{{\rm OHF} \rightarrow e} = \frac{A_{N} - f_{h^{\pm}}A_{N}^{h^{\pm}} - f_{J/\psi }A_{N}^{J/\psi}}{1 - f_{h^{\pm}} - f_{J/\psi }  - f_{\pi^{0} } - f_{\eta } - f_{\gamma }}
\label{eqn:BackgroundCorrection},
\end{equation}
where $f_{i}$ represent the background fractions, $A_{N}$ is the 
asymmetry calculated on the electron candidate sample, and $A_{N}^{i}$ 
are the background asymmetries. The procedure to calculate the 
background fractions and a more detailed description of background 
sources can be found in Ref.~\cite{PPG223}. This procedure is repeated 
in this analysis with the relevant $p_{T}$ bins, and 
uncertainties on calculated background fractions are propagated through 
Eq.~(\ref{eqn:BackgroundCorrection}) to obtain systematic uncertainties 
$\sigma^{\rm sys}_{f^{\pm}}$. Figure~\ref{fig:BGFractionPlot} shows the 
resulting background fractions for electrons and positrons combined. 
The Ke3 background source, which consists of Dalitz 
decays of $K^{\pm}$ and $K^{0}_{S}$, is heavily suppressed over the 
measured $p_{T}$ range. The transverse single-spin asymmetries for 
$K^{\pm}$ or $K^{0}_{S}$ have not been measured in $\sqrt{s} = 200$ GeV 
$p^{\uparrow}+p$ collisions. However, given that the Ke3 background 
fraction is on the order of $10^{-3}$, and is the smallest 
contributor, it is safely neglected in the background correction 
procedure. The relevant background fractions are calculated separately 
for positrons and electrons as shown in Table~\ref{tab:bgFracs}, with 
resulting background fractions shown in 
Figures~\ref{fig:BGFractionPlotp} and~\ref{fig:BGFractionPlotm}.

\begingroup \squeezetable
    \begin{table*}[htp] 
        \caption{Fractions of background $f_{i}$ present in each $p_{T}$ 
bin for the open-heavy-flavor positrons and electrons, used as inputs to 
the background correction procedure, and shown in 
Figs.~\ref{fig:BGFractionPlotp} and~\ref{fig:BGFractionPlotm} 
respectively.}
            \begin{ruledtabular}\begin{tabular}{cccccccc}
                $e^{\pm}$ & $p_{T}$ range (GeV/c) & $\left< p_{T} \right>$ (GeV/c) & $f_{\pi^{0} \rightarrow e^{\pm}}$ & $f_{\eta \rightarrow e^{\pm}}$ & $f_{\gamma \rightarrow e^{\pm}}$ & $f_{J/\psi \rightarrow e^{\pm}}$ & $f_{h^{\pm}}$ \\
                \hline
                $e^+$ &  1.0 -- 1.3 & 1.161  & 0.458 & 0.0738 & 0.00274 & 0.00916 & 0.0140 \\ 
                & 1.3 -- 1.5 & 1.398  & 0.318 & 0.0592 & 0.00336 & 0.0195 & 0.00924 \\ 
                & 1.5 -- 1.8 & 1.639  & 0.264 & 0.0582 & 0.00339 & 0.0344 & 0.0120 \\ 
                & 1.8 -- 2.1 & 1.936  & 0.215 & 0.0458 & 0.00399 & 0.0520 & 0.0134 \\ 
                & 2.1 -- 2.7 & 2.349  & 0.173 & 0.0394 & 0.00481 & 0.0823 & 0.0179 \\ 
                &2.7 -- 5.0 & 3.290  & 0.111 & 0.0297 & 0.00480 & 0.122 & 0.0300 \\ 
                \\
             
                $e^{-}$ & 1.0 -- 1.3 & 1.161  & 0.439 & 0.0704 & 0.00335 & 0.00900 & 0.0261 \\ 
                & 1.3 -- 1.5 & 1.398  & 0.347 & 0.0692 & 0.00364 & 0.0206 & 0.0198 \\ 
                & 1.5 -- 1.8 & 1.639  & 0.299 & 0.0665 & 0.00394 & 0.0375 & 0.0230 \\ 
                & 1.8 -- 2.1 & 1.936  & 0.252 & 0.0478 & 0.00535 & 0.0577 & 0.0205 \\ 
                & 2.1 -- 2.7 & 2.349  & 0.208 & 0.0429 & 0.00490 & 0.0872 & 0.0245 \\ 
                & 2.7 -- 5.0 & 3.290  & 0.143 & 0.0296 & 0.00572 & 0.127 & 0.0279 \\              
            \end{tabular}\end{ruledtabular}
        \label{tab:bgFracs}
    \end{table*} 
\endgroup

TSSAs for each background source have been measured at PHENIX at 
midrapidity in $p^{\uparrow}+p$ collisions at $\sqrt{s}=200$~GeV.  The 
asymmetries for photonic background sources $\pi^{0}, \eta$, and 
$\gamma$ were all measured by PHENIX to be consistent with zero using 
the same dataset as this measurement~\cite{PPG234,PPG235}. They are 
therefore set to zero in Eq.~(\ref{eqn:BackgroundCorrection}), with a 
systematic uncertainty $\sigma^{\rm sys}_{A_{N}^{B}}$ assigned for 
setting $A_{N}^{\pi^{0}}=0$ and $A_{N}^{\eta}=0$ based on propagating 
uncertainties from the measurements, while the uncertainty associated 
with setting the direct photon TSSA $A_{N}^{\gamma}=0$ is negligible 
because $f_{\gamma}$ is on the order of $10^{-3}$ (see 
Table~\ref{tab:bgFracs}). The TSSAs for $J/\psi$~\cite{PPG103E} and 
charged hadrons~\cite{PPG050} were measured with previous PHENIX data 
sets. The TSSA for $J/\psi$ has a large statistical 
uncertainty~\cite{PPG103E}, and contributes significantly to the 
statistical uncertainty of this measurement, especially at high $p_{T}$. 
This is due to the azimuthal angle of the decay lepton becoming more 
strongly correlated with the azimuthal angle of the $J/\psi$ at higher 
$p_{T}$. Additionally, the statistical precision of $A_{N} (p^{\uparrow} 
+ p \rightarrow J/\psi + X)$ could not be improved upon in the Run-15 
data given the high degree of photonic electron background. The TSSA for 
midrapidity $J/\psi$ production measured in Ref.~\cite{PPG103E} was 
recalculated as a function of decay lepton $p_{T}$ using {\sc 
pythia}~\cite{Sjostrand:2000wi} decay simulations for the $J/\psi 
\rightarrow e^{+}e^{-}$ channel to apply 
Eq.~(\ref{eqn:BackgroundCorrection}).

Due to the large contribution of statistical uncertainty coming from 
propagating the previously measured $A_{N} 
(p^{\uparrow}+p{\rightarrow}J/\psi+X)$ from Ref.~\cite{PPG103E} 
through the background correction formula 
(Eq.~\ref{eqn:BackgroundCorrection}), we report nonphotonic electron 
and positron asymmetries in addition to the open-heavy-flavor-decay 
electron and positron asymmetries. This allows the statistical 
precision of the open-heavy-flavor result to be improved upon given a 
more statistically precise measurement of 
$A_{N}(p^{\uparrow}+p{\rightarrow}J/\psi+X)$. 
Figures~\ref{fig:finalANs} and ~\ref{fig:LamParam} do not show the 
nonphotonic electron asymmetries because they are not the focus of 
this paper.  However, these asymmetries are shown and discussed 
below.  The formula for extracting the nonphotonic electron (${\rm 
NP}e$) asymmetry from the electron candidate sample asymmetry is
\begin{equation}
    A_{N}^{{\rm NP}e} = \frac{A_{N} - f_{h^{\pm}}A_{N}^{h^{\pm}} }{1 - f_{h^{\pm}}  - f_{\pi^{0} } - f_{\eta } - f_{\gamma }}
\label{eqn:BackgroundCorrectionNP}.
\end{equation}
\noindent Note that Eq.~\ref{eqn:BackgroundCorrectionNP} only differs 
from Eq.~\ref{eqn:BackgroundCorrection} by the omission of the terms 
including $J/\psi$ background fractions and asymmetries.

The TSSAs for midrapidity open charm production ($A_{N}^{D}$) predicted 
in Refs.~\cite{trigluon_twists} and~\cite{trigluon_twists_Yoshida} were 
also recalculated as a function of decay lepton $p_{T}$ for all possible 
semileptonic decay channels, with decay kinematics simulated in {\sc 
pythia}~\cite{Sjostrand:2000wi} to obtain correlations between $p_{T}$ 
and $\phi$ of the decay lepton and $D$ meson. The $\phi^{e}$ 
distribution was then weighted in accordance with 
$w(\phi^{e})=1+A_{N}^D(p_{T}^D)\cos\phi^{D}$ in each $p_{T}$ bin and 
then fit with $f(\phi) = N_{0}(1+A_{N}^{e} \cos \phi)$ to extract the 
decay lepton asymmetry. $D^{0}$ and $\bar{D}^{0}$ production was 
considered for comparisons to results from Refs.~\cite{trigluon_twists} 
and~\cite{trigluon_twists_Yoshida}, while $D^{+}$ and $D^{-}$ production 
was additionally considered when comparing to results of 
Ref.~\cite{trigluon_twists_Yoshida}. OHF production is dominated by open 
charm at the relevant kinematics, for which $D^{0}, \bar{D}^{0}, D^{+}$, 
and $D^{-}$ cover a significant fraction. The effect of including 
$D^{+}$ and $D^{-}$ in comparing to Ref.~\cite{trigluon_twists_Yoshida} 
makes very little difference as supported by our simulations, implying 
that comparing to $D^{0}$ and $\bar{D}^{0}$ for 
Ref.~\cite{trigluon_twists} is sufficient. A scan over $(\lambda_{f}, 
\lambda_{d})$ parameter space and independent scans over $K_{G}$ and 
$K_{G}'$ were performed to generate a set of theory curves for 
comparison, allowing for best-fit parameters and confidence intervals to 
be determined from data.


\begingroup \squeezetable
\begin{table*}[htb]
    \caption{Summary of final asymmetries $A_N^{{\rm 
OHF}\rightarrow{e^{\pm}}}$ for open-heavy-flavor positrons and electrons 
with statistical $\sigma^{A_N^{{\rm OHF}\rightarrow{e^{\pm}}}}$ and 
systematic uncertainties, shown in Fig.~\ref{fig:finalANs}.}
\begin{ruledtabular} \begin{tabular}{ccccccccccc}
$e^\pm$ & $p_{T}$ range & $\left< p_{T} \right>$ &&&&&&&& \\
& (GeV/$c$)      & (GeV/$c$) & $A_N^{\rm OHF}\rightarrow{e^\pm}$ & $\sigma^{A_N^{\rm OHF}\rightarrow{e^\pm}}$ 
& $\sigma^{\rm sys}_{f^+}$ & $\sigma^{\rm sys}_{f^-}$ & $\sigma^{\rm sys}_{A_{N}^{B}}$ & $\sigma^{\rm sys}_{\rm diff}$ 
& $\sigma^{\rm sys}_{{\rm tot}^+}$ & $\sigma^{\rm sys}_{{\rm tot}^-}$ \\ 
            \hline
$e^+$ & 1.0--1.3 & 1.161  & -0.00256 & 0.0212 & 0.00193 & 0.000855 & 0.00264 & 0.000435 & 0.00330 & 0.00281 \\ 
     & 1.3--1.5 & 1.398  & 0.0105 & 0.0178 & 0.00142 & 0.00108 & 0.00143 & 0.000621 & 0.00211 & 0.00189 \\ 
     & 1.5--1.8 & 1.639  & 0.00571 & 0.0159 & 0.000468 & 0.000401 & 0.00118 & 0.000432 & 0.00134 & 0.00132 \\ 
     & 1.8--2.1 & 1.936  & 0.0126 & 0.0192 & 0.00101 & 0.000856 & 0.000889 & 0.00697 & 0.00710 & 0.00708 \\ 
     & 2.1--2.7 & 2.349  & 0.00208 & 0.0210 & 0.00140 & 0.00109 & 0.000719 & 0.00446 & 0.00473 & 0.00465 \\ 
     & 2.7--5.0 & 3.290  & 0.0357 & 0.0287 & 0.00595 & 0.00364 & 0.000474 & 0.00342 & 0.00688 & 0.00501 \\ 
\\
$e^-$& 1.0--1.3 & 1.161  & -0.0113 & 0.0186 & 0.00404 & 0.00237 & 0.00247 & 0.000120 & 0.00474 & 0.00343 \\ 
    &   1.3--1.5 & 1.398  & -0.0297 & 0.0181 & 0.00466 & 0.00335 & 0.00174 & 0.000672 & 0.00502 & 0.00384 \\ 
    &   1.5--1.8 & 1.639  & 0.0139 & 0.0167 & 0.00117 & 0.000789 & 0.00147 & 0.000917 & 0.00209 & 0.00191 \\ 
    &   1.8--2.1 & 1.936  & 0.0105 & 0.0207 & 0.00136 & 0.000990 & 0.00109 & 0.000234 & 0.00176 & 0.00149 \\ 
    &   2.1--2.7 & 2.349  & -0.0267 & 0.0227 & 0.000104 & 0.000152 & 0.000899 & 0.00253 & 0.00269 & 0.00269 \\ 
    &   2.7--5.0 & 3.290  & 0.0237 & 0.0305 & 0.00509 & 0.00313 & 0.000589 & 0.00174 & 0.00541 & 0.00363 \\ 
        \end{tabular}  \end{ruledtabular}
        \label{tab:finalANHF}
    \caption{Summary of final asymmetries $A_{N}^{\rm NP}e$ for 
nonphotonic positrons and electrons with statistical $\sigma^{A_N^{{\rm 
NP}e}}$ and systematic uncertainties.}
    \begin{ruledtabular}  \begin{tabular}{ccccccccccc}
$e^\pm$ &  $p_{T}$ range & $\left<p_{T}\right>$ & & & & & & & &  \\
        &  (GeV/$c$) & (GeV/$c$) & $A_{N}^{\rm NP}e$ & $\sigma^{A_{N}^{\rm NP}e}$ & $\sigma^{\rm sys}_{f^{+}}$ 
& $\sigma^{\rm sys}_{f^{-}}$ & $\sigma^{\rm sys}_{A_N^B}$ & $\sigma^{\rm sys}_{\rm diff}$ 
& $\sigma^{\rm sys}_{{\rm tot}^+}$ & $\sigma^{\rm sys}_{{\rm tot}^-}$ \\
        \hline
$e^+$ &  1.0-- 1.3 & 1.161  & -0.00202 & 0.0207 & 0.00115 & 0.000531 & 0.00259 & 0.000435 & 0.00286 & 0.00268 \\
      &  1.3-- 1.5 & 1.398  & 0.0103 & 0.0172 & 0.00128 & 0.000996 & 0.00138 & 0.000621 & 0.00198 & 0.00181 \\
      &  1.5-- 1.8 & 1.639  & 0.00379 & 0.0148 & 0.000119 & 8.15e-05 & 0.00112 & 0.000432 & 0.00120 & 0.00120 \\
      &  1.8-- 2.1 & 1.936  & 0.00836 & 0.0170 & 0.000222 & 0.000144 & 0.000825 & 0.00697 & 0.00702 & 0.00702 \\
      &  2.1-- 2.7 & 2.349  & -0.00371 & 0.0163 & 0.000239 & 7.51e-05 & 0.000642 & 0.00446 & 0.00452 & 0.00451 \\
      &  2.7-- 5.0 & 3.290  & 0.0220 & 0.0201 & 0.00205 & 0.000814 & 0.000404 & 0.00342 & 0.00401 & 0.00354 \\
\\
$e^-$ & 1.0--1.3 & 1.161  & -0.0106 & 0.0182 & 0.00338 & 0.00203 & 0.00242 & 0.000120 & 0.00416 & 0.00316 \\
      &  1.3--1.5 & 1.398  & -0.0284 & 0.0174 & 0.00386 & 0.00284 & 0.00168 & 0.000672 & 0.00426 & 0.00337 \\
      &  1.5--1.8 & 1.639  & 0.0111 & 0.0153 & 0.000538 & 0.000288 & 0.00138 & 0.000917 & 0.00174 & 0.00168 \\
      &  1.8--2.1 & 1.936  & 0.00565 & 0.0178 & 0.000282 & 0.000131 & 0.000996 & 0.000234 & 0.00106 & 0.00103 \\
      &  2.1--2.7 & 2.349  & -0.0297 & 0.0171 & 0.000446 & 0.000351 & 0.000790 & 0.00253 & 0.00269 & 0.00268 \\
      &  2.7--5.0 & 3.290  & 0.0108 & 0.0207 & 0.00134 & 0.000466 & 0.000495 & 0.00174 & 0.00225 & 0.00187 \\
	\end{tabular} \end{ruledtabular}
    \label{tab:finalANNP}
\end{table*}
\endgroup

\section{Results}

The ${\rm OHF}\rightarrow{e^{\pm}}$ TSSAs are plotted in 
Fig.~\ref{fig:finalANs} alongside theoretical predictions of 
$A_{N}(p^{\uparrow} + p \rightarrow (D^{0}/\bar{D^{0}}\rightarrow e^{\pm}) + X)$ from Ref.~\cite{trigluon_twists} in 
(red/blue) solid lines, and $A_{N}(p^{\uparrow} + p \rightarrow (D/\bar{D}\rightarrow e^{\pm}) + X)$ from 
Ref.~\cite{trigluon_twists_Yoshida} in (red/blue) dashed and dotted 
lines, with $\lambda_{f}$, $\lambda_{d}$, $K_{G}$ and $K_{G}'$ chosen to 
best fit the data for the separate charges simultaneously. The 
measurements are consistent with zero, and are statistically more 
precise than previous heavy-flavor measurements. The total systematic 
uncertainties come from combining those associated with the background 
fractions, background asymmetries, and the difference in calculating 
$A_{N}$ with Eqs.~(\ref{eqn:AN_lumi}) and~(\ref{eqn:AN_sqrt}); there is 
no dominant source of systematic uncertainty across charges and $p_{T}$ 
bins. The systematic uncertainty reaches at most 37\% of the 
corresponding statistical uncertainty (see Table~\ref{tab:finalANHF}), 
while it is typically suppressed by an order of magnitude or more. The 
placement of the theoretical curves in Fig.~\ref{fig:finalANs} differs 
for $e^{+}$ vs~$e^{-}$ due to the contribution of the symmetric trigluon 
correlator having opposing signs in \textit{charm} vs~\textit{anticharm} 
production, leading to constructive vs destructive interference with the 
antisymmetric trigluon correlator contribution for the separate charges. 
This allows for constraining power on all parameters. Summaries for 
final asymmetries with statistical and systematic uncertainties are 
given in Table~\ref{tab:finalANHF} for OHF positrons $A_N^{{\rm 
OHF}\rightarrow{e^+}}$ and electrons $A_N^{{\rm OHF}\rightarrow{e^-}}$ 
and in Table~\ref{tab:finalANNP} for nonphotonic (NP) positrons 
$A_N^{{\rm NP}{e^+}}$ and electrons $A_N^{{\rm NP}{e^-}}$.

To determine theoretical parameters that fit the data best, 
$\chi^{2}(\lambda_{f},\lambda_{d})$, $\chi^{2}(K_{G})$, and 
$\chi^{2}(K_{G}')$ were calculated for the separate charges and summed 
to extract minimum values. The results along with $1 \sigma$ confidence 
intervals are $\lambda_{f}=-0.01{\pm}0.03$ GeV and 
$\lambda_{d}=0.11{\pm}0.09$ GeV for parameters introduced 
Ref.~\cite{trigluon_twists}, and $K_{G}=0.0006^{+0.0014}_{-0.0017}$, and 
$K_{G}'=0.00025{\pm}0.00022$ for parameters introduced in 
Ref.~\cite{trigluon_twists_Yoshida}. This corresponds to the first 
constraints on $(\lambda_{f}, \lambda_{d})$, and is in agreement with 
previous constraints on $K_{G}$ and $K_{G}'$ derived in 
Ref.~\cite{trigluon_twists_Yoshida}.  Figure~\ref{fig:LamParam} 
summarizes the results of the statistical analysis performed to extract 
best-fit parameters $\lambda_{f}$ and $\lambda_{d}$, where the 
theoretical asymmetries depend on both parameters. Nicely illustrated 
are the constraining power of the individual charges and the necessity 
of combining the charges in the statistical analysis. Both charges 
predict that contributions from trigluon correlations are small, 
indicating that $\lambda_{f}$ and $\lambda_{d}$ values that result in 
cancellation of their contributions to the asymmetry calculation are 
preferred.

\section{Summary}

In summary, the PHENIX experiment has measured the transverse 
single-spin asymmetry of midrapidity open-heavy-flavor decay electrons 
and positrons as a function of $p_{T}$ in $p^{\uparrow}+p$ collisions at 
$\sqrt{s}=200$~GeV. Open-heavy-flavor production at RHIC is an ideal 
channel for probing trigluon correlations in polarized protons because 
initial-state $qgq$ correlations in the proton and final-state twist-3 
correlations in hadronization contribute negligibly. This measurement 
provides constraints for the antisymmetric and symmetric trigluon 
correlation functions in transversely polarized protons, including the 
first constraints on $\lambda_{f}$ and $\lambda_{d}$ as 
$\lambda_f=-0.01{\pm}0.03$~GeV and $\lambda_d=0.11{\pm}0.09$~GeV ---  a 
necessary step forward in our understanding of proton structure through 
correlations between proton spin and gluon momentum.



\begin{acknowledgments}

We thank the staff of the Collider-Accelerator and Physics
Departments at Brookhaven National Laboratory and the staff of
the other PHENIX participating institutions for their vital
contributions.  
We also thank Z. Kang and S. Yoshida for providing $A_{N}^{D}(p_{T})$ 
calculations corresponding to the models introduced in 
Refs.~\cite{trigluon_twists} and~\cite{trigluon_twists_Yoshida}.
We acknowledge support from the Office of Nuclear Physics in the
Office of Science of the Department of Energy,
the National Science Foundation,
Abilene Christian University Research Council,
Research Foundation of SUNY, and
Dean of the College of Arts and Sciences, Vanderbilt University
(U.S.A),
Ministry of Education, Culture, Sports, Science, and Technology
and the Japan Society for the Promotion of Science (Japan),
Natural Science Foundation of China (People's Republic of China),
Croatian Science Foundation and
Ministry of Science and Education (Croatia),
Ministry of Education, Youth and Sports (Czech Republic),
Centre National de la Recherche Scientifique, Commissariat
{\`a} l'{\'E}nergie Atomique, and Institut National de Physique
Nucl{\'e}aire et de Physique des Particules (France),
J. Bolyai Research Scholarship, EFOP, the New National Excellence
Program ({\'U}NKP), NKFIH, and OTKA (Hungary),
Department of Atomic Energy and Department of Science and Technology
(India),
Israel Science Foundation (Israel),
Basic Science Research and SRC(CENuM) Programs through NRF
funded by the Ministry of Education and the Ministry of
Science and ICT (Korea).
Ministry of Education and Science, Russian Academy of Sciences,
Federal Agency of Atomic Energy (Russia),
VR and Wallenberg Foundation (Sweden),
University of Zambia, the Government of the Republic of Zambia (Zambia),
the U.S. Civilian Research and Development Foundation for the
Independent States of the Former Soviet Union,
the Hungarian American Enterprise Scholarship Fund,
the US-Hungarian Fulbright Foundation,
and the US-Israel Binational Science Foundation.

\end{acknowledgments}




\begin{thebibliography}{54}%
\makeatletter
\providecommand \@ifxundefined [1]{%
 \@ifx{#1\undefined}
}%
\providecommand \@ifnum [1]{%
 \ifnum #1\expandafter \@firstoftwo
 \else \expandafter \@secondoftwo
 \fi
}%
\providecommand \@ifx [1]{%
 \ifx #1\expandafter \@firstoftwo
 \else \expandafter \@secondoftwo
 \fi
}%
\providecommand \natexlab [1]{#1}%
\providecommand \enquote  [1]{``#1''}%
\providecommand \bibnamefont  [1]{#1}%
\providecommand \bibfnamefont [1]{#1}%
\providecommand \citenamefont [1]{#1}%
\providecommand \href@noop [0]{\@secondoftwo}%
\providecommand \href [0]{\begingroup \@sanitize@url \@href}%
\providecommand \@href[1]{\@@startlink{#1}\@@href}%
\providecommand \@@href[1]{\endgroup#1\@@endlink}%
\providecommand \@sanitize@url [0]{\catcode `\\12\catcode `\$12\catcode
  `\&12\catcode `\#12\catcode `\^12\catcode `\_12\catcode `\%12\relax}%
\providecommand \@@startlink[1]{}%
\providecommand \@@endlink[0]{}%
\providecommand \url  [0]{\begingroup\@sanitize@url \@url }%
\providecommand \@url [1]{\endgroup\@href {#1}{\urlprefix }}%
\providecommand \urlprefix  [0]{URL }%
\providecommand \Eprint [0]{\href }%
\providecommand \doibase [0]{https://doi.org/}%
\providecommand \selectlanguage [0]{\@gobble}%
\providecommand \bibinfo  [0]{\@secondoftwo}%
\providecommand \bibfield  [0]{\@secondoftwo}%
\providecommand \translation [1]{[#1]}%
\providecommand \BibitemOpen [0]{}%
\providecommand \bibitemStop [0]{}%
\providecommand \bibitemNoStop [0]{.\EOS\space}%
\providecommand \EOS [0]{\spacefactor3000\relax}%
\providecommand \BibitemShut  [1]{\csname bibitem#1\endcsname}%
\let\auto@bib@innerbib\@empty
\bibitem [{\citenamefont {Klem}\ \emph {et~al.}(1976)\citenamefont {Klem},
  \citenamefont {Bowers}, \citenamefont {Courant}, \citenamefont {Kagan},
  \citenamefont {Marshak}, \citenamefont {Peterson}, \citenamefont {Ruddick},
  \citenamefont {Dragoset},\ and\ \citenamefont {Roberts}}]{Klem:1976ui}%
  \BibitemOpen
  \bibfield  {author} {\bibinfo {author} {\bibfnamefont {R.~D.}\ \bibnamefont
  {Klem}}, \bibinfo {author} {\bibfnamefont {J.~E.}\ \bibnamefont {Bowers}},
  \bibinfo {author} {\bibfnamefont {H.~W.}\ \bibnamefont {Courant}}, \bibinfo
  {author} {\bibfnamefont {H.}~\bibnamefont {Kagan}}, \bibinfo {author}
  {\bibfnamefont {M.~L.}\ \bibnamefont {Marshak}}, \bibinfo {author}
  {\bibfnamefont {E.~A.}\ \bibnamefont {Peterson}}, \bibinfo {author}
  {\bibfnamefont {K.}~\bibnamefont {Ruddick}}, \bibinfo {author} {\bibfnamefont
  {W.~H.}\ \bibnamefont {Dragoset}},\ and\ \bibinfo {author} {\bibfnamefont
  {J.~B.}\ \bibnamefont {Roberts}},\ }\bibfield  {title} {\bibinfo {title}
  {{Measurement of Asymmetries of Inclusive Pion Production in Proton Proton
  Interactions at 6 GeV/$c$ and 11.8 GeV/$c$}},\ }\href
  {https://doi.org/10.1103/PhysRevLett.36.929} {\bibfield  {journal} {\bibinfo
  {journal} {Phys. Rev. Lett.}\ }\textbf {\bibinfo {volume} {36}},\ \bibinfo
  {pages} {929} (\bibinfo {year} {1976})}\BibitemShut {NoStop}%
\bibitem [{\citenamefont {Adams}\ \emph {et~al.}(1991)\citenamefont {Adams}
  \emph {et~al.}}]{Adams:1991cs}%
  \BibitemOpen
  \bibfield  {author} {\bibinfo {author} {\bibfnamefont {D.~L.}\ \bibnamefont
  {Adams}} \emph {et~al.} (\bibinfo {collaboration} {FNAL-E704
  Collaboration}),\ }\bibfield  {title} {\bibinfo {title} {{Analyzing power in
  inclusive $\pi^+$ and $\pi^-$ production at high $x_F$ with a 200-GeV
  polarized proton beam}},\ }\href
  {https://doi.org/10.1016/0370-2693(91)90378-4} {\bibfield  {journal}
  {\bibinfo  {journal} {Phys. Lett. B}\ }\textbf {\bibinfo {volume} {264}},\
  \bibinfo {pages} {462} (\bibinfo {year} {1991})}\BibitemShut {NoStop}%
\bibitem [{\citenamefont {Allgower}\ \emph {et~al.}(2002)\citenamefont
  {Allgower} \emph {et~al.}}]{Allgower:2002qi}%
  \BibitemOpen
  \bibfield  {author} {\bibinfo {author} {\bibfnamefont {C.~E.}\ \bibnamefont
  {Allgower}} \emph {et~al.},\ }\bibfield  {title} {\bibinfo {title}
  {{Measurement of analyzing powers of $\pi^+$ and $\pi^-$ produced on a
  hydrogen and a carbon target with a 22-GeV/$c$ incident polarized proton
  beam}},\ }\href {https://doi.org/10.1103/PhysRevD.65.092008} {\bibfield
  {journal} {\bibinfo  {journal} {Phys. Rev. D}\ }\textbf {\bibinfo {volume}
  {65}},\ \bibinfo {pages} {092008} (\bibinfo {year} {2002})}\BibitemShut
  {NoStop}%
\bibitem [{\citenamefont {Arsene}\ \emph {et~al.}(2008)\citenamefont {Arsene}
  \emph {et~al.}}]{Arsene:2008aa}%
  \BibitemOpen
  \bibfield  {author} {\bibinfo {author} {\bibfnamefont {I.}~\bibnamefont
  {Arsene}} \emph {et~al.} (\bibinfo {collaboration} {BRAHMS Collaboration}),\
  }\bibfield  {title} {\bibinfo {title} {{Single Transverse Spin Asymmetries of
  Identified Charged Hadrons in Polarized $p+p$ Collisions at $\sqrt{s}$ = 62.4
  GeV}},\ }\href {https://doi.org/10.1103/PhysRevLett.101.042001} {\bibfield
  {journal} {\bibinfo  {journal} {Phys. Rev. Lett.}\ }\textbf {\bibinfo
  {volume} {101}},\ \bibinfo {pages} {042001} (\bibinfo {year}
  {2008})}\BibitemShut {NoStop}%
\bibitem [{\citenamefont {Kane}\ \emph {et~al.}(1978)\citenamefont {Kane},
  \citenamefont {Pumplin},\ and\ \citenamefont {Repko}}]{pQCDTSSA}%
  \BibitemOpen
  \bibfield  {author} {\bibinfo {author} {\bibfnamefont {G.~L.}\ \bibnamefont
  {Kane}}, \bibinfo {author} {\bibfnamefont {J.}~\bibnamefont {Pumplin}},\ and\
  \bibinfo {author} {\bibfnamefont {W.}~\bibnamefont {Repko}},\ }\bibfield
  {title} {\bibinfo {title} {{Transverse Quark Polarization in Large-${p}_{T}$
  Reactions, $e^+e^-$ Jets, and Leptoproduction: A Test of Quantum
  Chromodynamics}},\ }\href {https://doi.org/10.1103/PhysRevLett.41.1689}
  {\bibfield  {journal} {\bibinfo  {journal} {Phys. Rev. Lett.}\ }\textbf
  {\bibinfo {volume} {41}},\ \bibinfo {pages} {1689} (\bibinfo {year}
  {1978})}\BibitemShut {NoStop}%
\bibitem [{\citenamefont {Benic}\ \emph {et~al.}(2019)\citenamefont {Benic},
  \citenamefont {Hatta}, \citenamefont {Li},\ and\ \citenamefont
  {Yang}}]{twoLoopsTSSA}%
  \BibitemOpen
  \bibfield  {author} {\bibinfo {author} {\bibfnamefont {S.}~\bibnamefont
  {Benic}}, \bibinfo {author} {\bibfnamefont {Y.}~\bibnamefont {Hatta}},
  \bibinfo {author} {\bibfnamefont {H.-n.}\ \bibnamefont {Li}},\ and\ \bibinfo
  {author} {\bibfnamefont {D.-J.}\ \bibnamefont {Yang}},\ }\bibfield  {title}
  {\bibinfo {title} {{Single-spin asymmetries at two loops}},\ }\href
  {https://doi.org/10.1103/PhysRevD.100.094027} {\bibfield  {journal} {\bibinfo
   {journal} {Phys. Rev. D}\ }\textbf {\bibinfo {volume} {100}},\ \bibinfo
  {pages} {094027} (\bibinfo {year} {2019})}\BibitemShut {NoStop}%
\bibitem [{\citenamefont {Sivers}(1990)}]{SiversTMD}%
  \BibitemOpen
  \bibfield  {author} {\bibinfo {author} {\bibfnamefont {D.}~\bibnamefont
  {Sivers}},\ }\bibfield  {title} {\bibinfo {title} {{Single-spin production
  asymmetries from the hard scattering of pointlike constituents}},\ }\href
  {https://doi.org/10.1103/PhysRevD.41.83} {\bibfield  {journal} {\bibinfo
  {journal} {Phys. Rev. D}\ }\textbf {\bibinfo {volume} {41}},\ \bibinfo
  {pages} {83} (\bibinfo {year} {1990})}\BibitemShut {NoStop}%
\bibitem [{\citenamefont {Boer}\ and\ \citenamefont
  {Mulders}(1998)}]{BoerMuldersTMD}%
  \BibitemOpen
  \bibfield  {author} {\bibinfo {author} {\bibfnamefont {D.}~\bibnamefont
  {Boer}}\ and\ \bibinfo {author} {\bibfnamefont {P.~J.}\ \bibnamefont
  {Mulders}},\ }\bibfield  {title} {\bibinfo {title} {{Time-reversal odd
  distribution functions in leptoproduction}},\ }\href
  {https://doi.org/10.1103/PhysRevD.57.5780} {\bibfield  {journal} {\bibinfo
  {journal} {Phys. Rev. D}\ }\textbf {\bibinfo {volume} {57}},\ \bibinfo
  {pages} {5780} (\bibinfo {year} {1998})}\BibitemShut {NoStop}%
\bibitem [{\citenamefont {Collins}(1993)}]{CollinsTMDFF}%
  \BibitemOpen
  \bibfield  {author} {\bibinfo {author} {\bibfnamefont {J.}~\bibnamefont
  {Collins}},\ }\bibfield  {title} {\bibinfo {title} {{Fragmentation of
  transversely polarized quarks probed in transverse momentum distributions}},\
  }\href {https://doi.org/https://doi.org/10.1016/0550-3213(93)90262-N}
  {\bibfield  {journal} {\bibinfo  {journal} {Nucl. Phys. B}\ }\textbf
  {\bibinfo {volume} {396}},\ \bibinfo {pages} {161} (\bibinfo {year}
  {1993})}\BibitemShut {NoStop}%
\bibitem [{\citenamefont {Efremov}\ and\ \citenamefont
  {Teryaev}(1985)}]{Efremov:1984ip}%
  \BibitemOpen
  \bibfield  {author} {\bibinfo {author} {\bibfnamefont {A.~V.}\ \bibnamefont
  {Efremov}}\ and\ \bibinfo {author} {\bibfnamefont {O.~V.}\ \bibnamefont
  {Teryaev}},\ }\bibfield  {title} {\bibinfo {title} {{QCD Asymmetry and
  Polarized Hadron Structure Functions}},\ }\href
  {https://doi.org/10.1016/0370-2693(85)90999-2} {\bibfield  {journal}
  {\bibinfo  {journal} {Phys. Lett. B}\ }\textbf {\bibinfo {volume} {150}},\
  \bibinfo {pages} {383} (\bibinfo {year} {1985})}\BibitemShut {NoStop}%
\bibitem [{\citenamefont {Qiu}\ and\ \citenamefont
  {Sterman}(1991)}]{SSA_twist3}%
  \BibitemOpen
  \bibfield  {author} {\bibinfo {author} {\bibfnamefont {J.-W.}\ \bibnamefont
  {Qiu}}\ and\ \bibinfo {author} {\bibfnamefont {G.}~\bibnamefont {Sterman}},\
  }\bibfield  {title} {\bibinfo {title} {Single transverse spin asymmetries},\
  }\href {https://doi.org/10.1103/PhysRevLett.67.2264} {\bibfield  {journal}
  {\bibinfo  {journal} {Phys. Rev. Lett.}\ }\textbf {\bibinfo {volume} {67}},\
  \bibinfo {pages} {2264} (\bibinfo {year} {1991})}\BibitemShut {NoStop}%
\bibitem [{\citenamefont {Anselmino}\ \emph {et~al.}(2020)\citenamefont
  {Anselmino}, \citenamefont {Mukherjee},\ and\ \citenamefont
  {Vossen}}]{TSSA_review}%
  \BibitemOpen
  \bibfield  {author} {\bibinfo {author} {\bibfnamefont {M.}~\bibnamefont
  {Anselmino}}, \bibinfo {author} {\bibfnamefont {A.}~\bibnamefont
  {Mukherjee}},\ and\ \bibinfo {author} {\bibfnamefont {A.}~\bibnamefont
  {Vossen}},\ }\bibfield  {title} {\bibinfo {title} {{Transverse spin effects
  in hard semi-inclusive collisions}},\ }\href
  {https://doi.org/10.1016/j.ppnp.2020.103806} {\bibfield  {journal} {\bibinfo
  {journal} {Prog. Part. Nucl. Phys.}\ }\textbf {\bibinfo {volume} {114}},\
  \bibinfo {pages} {103806} (\bibinfo {year} {2020})}\BibitemShut {NoStop}%
\bibitem [{\citenamefont {Cammarota}\ \emph {et~al.}(2020)\citenamefont
  {Cammarota}, \citenamefont {Gamberg}, \citenamefont {Kang}, \citenamefont
  {Miller}, \citenamefont {Pitonyak}, \citenamefont {Prokudin}, \citenamefont
  {Rogers},\ and\ \citenamefont {Sato}}]{SSAorigin}%
  \BibitemOpen
  \bibfield  {author} {\bibinfo {author} {\bibfnamefont {J.}~\bibnamefont
  {Cammarota}}, \bibinfo {author} {\bibfnamefont {L.}~\bibnamefont {Gamberg}},
  \bibinfo {author} {\bibfnamefont {Z.-B.}\ \bibnamefont {Kang}}, \bibinfo
  {author} {\bibfnamefont {J.~A.}\ \bibnamefont {Miller}}, \bibinfo {author}
  {\bibfnamefont {D.}~\bibnamefont {Pitonyak}}, \bibinfo {author}
  {\bibfnamefont {A.}~\bibnamefont {Prokudin}}, \bibinfo {author}
  {\bibfnamefont {T.~C.}\ \bibnamefont {Rogers}},\ and\ \bibinfo {author}
  {\bibfnamefont {N.}~\bibnamefont {Sato}} (\bibinfo {collaboration} {Jefferson
  Lab Angular Momentum Collaboration}),\ }\bibfield  {title} {\bibinfo {title}
  {{Origin of single transverse-spin asymmetries in high-energy collisions}},\
  }\href {https://doi.org/10.1103/PhysRevD.102.054002} {\bibfield  {journal}
  {\bibinfo  {journal} {Phys. Rev. D}\ }\textbf {\bibinfo {volume} {102}},\
  \bibinfo {pages} {054002} (\bibinfo {year} {2020})}\BibitemShut {NoStop}%
\bibitem [{\citenamefont {Ji}\ \emph {et~al.}(2006)\citenamefont {Ji},
  \citenamefont {Qiu}, \citenamefont {Vogelsang},\ and\ \citenamefont
  {Yuan}}]{Ji:2006ub}%
  \BibitemOpen
  \bibfield  {author} {\bibinfo {author} {\bibfnamefont {X.}~\bibnamefont
  {Ji}}, \bibinfo {author} {\bibfnamefont {J.-W.}\ \bibnamefont {Qiu}},
  \bibinfo {author} {\bibfnamefont {W.}~\bibnamefont {Vogelsang}},\ and\
  \bibinfo {author} {\bibfnamefont {F.}~\bibnamefont {Yuan}},\ }\bibfield
  {title} {\bibinfo {title} {{A Unified picture for single transverse-spin
  asymmetries in hard processes}},\ }\href
  {https://doi.org/10.1103/PhysRevLett.97.082002} {\bibfield  {journal}
  {\bibinfo  {journal} {Phys. Rev. Lett.}\ }\textbf {\bibinfo {volume} {97}},\
  \bibinfo {pages} {082002} (\bibinfo {year} {2006})}\BibitemShut {NoStop}%
\bibitem [{\citenamefont {Koike}\ \emph {et~al.}(2008)\citenamefont {Koike},
  \citenamefont {Vogelsang},\ and\ \citenamefont {Yuan}}]{TMD_twist3_compare}%
  \BibitemOpen
  \bibfield  {author} {\bibinfo {author} {\bibfnamefont {Y.}~\bibnamefont
  {Koike}}, \bibinfo {author} {\bibfnamefont {W.}~\bibnamefont {Vogelsang}},\
  and\ \bibinfo {author} {\bibfnamefont {F.}~\bibnamefont {Yuan}},\ }\bibfield
  {title} {\bibinfo {title} {On the relation between mechanisms for
  single-transverse-spin asymmetries},\ }\href
  {https://doi.org/https://doi.org/10.1016/j.physletb.2007.11.096} {\bibfield
  {journal} {\bibinfo  {journal} {Phys. Lett. B}\ }\textbf {\bibinfo {volume}
  {659}},\ \bibinfo {pages} {878} (\bibinfo {year} {2008})}\BibitemShut
  {NoStop}%
\bibitem [{\citenamefont {Yuan}\ and\ \citenamefont
  {Zhou}(2009)}]{Collins_twist3_compare}%
  \BibitemOpen
  \bibfield  {author} {\bibinfo {author} {\bibfnamefont {F.}~\bibnamefont
  {Yuan}}\ and\ \bibinfo {author} {\bibfnamefont {J.}~\bibnamefont {Zhou}},\
  }\bibfield  {title} {\bibinfo {title} {Collins function and the single
  transverse spin asymmetry},\ }\href
  {https://doi.org/10.1103/PhysRevLett.103.052001} {\bibfield  {journal}
  {\bibinfo  {journal} {Phys. Rev. Lett.}\ }\textbf {\bibinfo {volume} {103}},\
  \bibinfo {pages} {052001} (\bibinfo {year} {2009})}\BibitemShut {NoStop}%
\bibitem [{\citenamefont {Ji}\ \emph {et~al.}(2003)\citenamefont {Ji},
  \citenamefont {Ma},\ and\ \citenamefont {Yuan}}]{orbital1}%
  \BibitemOpen
  \bibfield  {author} {\bibinfo {author} {\bibfnamefont {X.-d.}\ \bibnamefont
  {Ji}}, \bibinfo {author} {\bibfnamefont {J.-P.}\ \bibnamefont {Ma}},\ and\
  \bibinfo {author} {\bibfnamefont {F.}~\bibnamefont {Yuan}},\ }\bibfield
  {title} {\bibinfo {title} {{Three quark light cone amplitudes of the proton
  and quark orbital motion dependent observables}},\ }\href@noop {} {\bibfield
  {journal} {\bibinfo  {journal} {Nucl. Phys. B}\ }\textbf {\bibinfo {volume}
  {652}},\ \bibinfo {pages} {383} (\bibinfo {year} {2003})}\BibitemShut
  {NoStop}%
\bibitem [{\citenamefont {Hatta}\ \emph {et~al.}()\citenamefont {Hatta},
  \citenamefont {Tanaka},\ and\ \citenamefont {Yoshida}}]{orbital2}%
  \BibitemOpen
  \bibfield  {author} {\bibinfo {author} {\bibfnamefont {Y.}~\bibnamefont
  {Hatta}}, \bibinfo {author} {\bibfnamefont {K.}~\bibnamefont {Tanaka}},\ and\
  \bibinfo {author} {\bibfnamefont {S.}~\bibnamefont {Yoshida}},\ }\href@noop
  {} {\bibinfo {title} {{Twist-three relations of gluonic correlators for the
  transversely polarized nucleon}}},\ \bibinfo {note} {{J. High Energy Phys. 02
  {\bf 04 (2013)} 003}}\BibitemShut {NoStop}%
\bibitem [{\citenamefont {Hatta}\ and\ \citenamefont {Yao}(2019)}]{orbital3}%
  \BibitemOpen
  \bibfield  {author} {\bibinfo {author} {\bibfnamefont {Y.}~\bibnamefont
  {Hatta}}\ and\ \bibinfo {author} {\bibfnamefont {X.}~\bibnamefont {Yao}},\
  }\bibfield  {title} {\bibinfo {title} {{QCD evolution of the orbital angular
  momentum of quarks and gluons: Genuine twist-three part}},\ }\href
  {https://doi.org/10.1016/j.physletb.2019.134941} {\bibfield  {journal}
  {\bibinfo  {journal} {Phys. Lett. B}\ }\textbf {\bibinfo {volume} {798}},\
  \bibinfo {pages} {134941} (\bibinfo {year} {2019})}\BibitemShut {NoStop}%
\bibitem [{\citenamefont {Qiu}\ and\ \citenamefont
  {Sterman}(1999)}]{TSSA_twist3_eqn_intro}%
  \BibitemOpen
  \bibfield  {author} {\bibinfo {author} {\bibfnamefont {J.-w.}\ \bibnamefont
  {Qiu}}\ and\ \bibinfo {author} {\bibfnamefont {G.~F.}\ \bibnamefont
  {Sterman}},\ }\bibfield  {title} {\bibinfo {title} {{Single transverse spin
  asymmetries in hadronic pion production}},\ }\href
  {https://doi.org/10.1103/PhysRevD.59.014004} {\bibfield  {journal} {\bibinfo
  {journal} {Phys. Rev. D}\ }\textbf {\bibinfo {volume} {59}},\ \bibinfo
  {pages} {014004} (\bibinfo {year} {1999})}\BibitemShut {NoStop}%
\bibitem [{\citenamefont {Kouvaris}\ \emph {et~al.}(2006)\citenamefont
  {Kouvaris}, \citenamefont {Qiu}, \citenamefont {Vogelsang},\ and\
  \citenamefont {Yuan}}]{TSSA_eqn}%
  \BibitemOpen
  \bibfield  {author} {\bibinfo {author} {\bibfnamefont {C.}~\bibnamefont
  {Kouvaris}}, \bibinfo {author} {\bibfnamefont {J.-W.}\ \bibnamefont {Qiu}},
  \bibinfo {author} {\bibfnamefont {W.}~\bibnamefont {Vogelsang}},\ and\
  \bibinfo {author} {\bibfnamefont {F.}~\bibnamefont {Yuan}},\ }\bibfield
  {title} {\bibinfo {title} {{Single transverse-spin asymmetry in high
  transverse momentum pion production in pp collisions}},\ }\href
  {https://doi.org/10.1103/PhysRevD.74.114013} {\bibfield  {journal} {\bibinfo
  {journal} {Phys. Rev. D}\ }\textbf {\bibinfo {volume} {74}},\ \bibinfo
  {pages} {114013} (\bibinfo {year} {2006})}\BibitemShut {NoStop}%
\bibitem [{\citenamefont {Ralston}\ and\ \citenamefont
  {Soper}(1979)}]{transversityDef}%
  \BibitemOpen
  \bibfield  {author} {\bibinfo {author} {\bibfnamefont {J.~P.}\ \bibnamefont
  {Ralston}}\ and\ \bibinfo {author} {\bibfnamefont {D.~E.}\ \bibnamefont
  {Soper}},\ }\bibfield  {title} {\bibinfo {title} {{Production of dimuons from
  high-energy polarized proton-proton collisions}},\ }\href
  {https://doi.org/https://doi.org/10.1016/0550-3213(79)90082-8} {\bibfield
  {journal} {\bibinfo  {journal} {Nucl. Phys. B}\ }\textbf {\bibinfo {volume}
  {152}},\ \bibinfo {pages} {109} (\bibinfo {year} {1979})}\BibitemShut
  {NoStop}%
\bibitem [{\citenamefont {Norrbin}\ and\ \citenamefont
  {Sjostrand}(2000)}]{heavy_quark_production}%
  \BibitemOpen
  \bibfield  {author} {\bibinfo {author} {\bibfnamefont {E.}~\bibnamefont
  {Norrbin}}\ and\ \bibinfo {author} {\bibfnamefont {T.}~\bibnamefont
  {Sjostrand}},\ }\bibfield  {title} {\bibinfo {title} {{Production and
  hadronization of heavy quarks}},\ }\href
  {https://doi.org/10.1007/s100520000460} {\bibfield  {journal} {\bibinfo
  {journal} {Eur. Phys. J. C}\ }\textbf {\bibinfo {volume} {17}},\ \bibinfo
  {pages} {137} (\bibinfo {year} {2000})}\BibitemShut {NoStop}%
\bibitem [{\citenamefont {Qiu}\ and\ \citenamefont
  {Sterman}(1992)}]{Qiu:1991wg}%
  \BibitemOpen
  \bibfield  {author} {\bibinfo {author} {\bibfnamefont {J.-W.}\ \bibnamefont
  {Qiu}}\ and\ \bibinfo {author} {\bibfnamefont {G.~F.}\ \bibnamefont
  {Sterman}},\ }\bibfield  {title} {\bibinfo {title} {{Single transverse spin
  asymmetries in direct photon production}},\ }\href
  {https://doi.org/10.1016/0550-3213(92)90003-T} {\bibfield  {journal}
  {\bibinfo  {journal} {Nucl. Phys. B}\ }\textbf {\bibinfo {volume} {378}},\
  \bibinfo {pages} {52} (\bibinfo {year} {1992})}\BibitemShut {NoStop}%
\bibitem [{\citenamefont {Ji}(1992)}]{trigluon_intro}%
  \BibitemOpen
  \bibfield  {author} {\bibinfo {author} {\bibfnamefont {X.}~\bibnamefont
  {Ji}},\ }\bibfield  {title} {\bibinfo {title} {{Gluon correlations in the
  transversely polarized nucleon}},\ }\href
  {https://doi.org/https://doi.org/10.1016/0370-2693(92)91375-J} {\bibfield
  {journal} {\bibinfo  {journal} {Phys. Lett. B}\ }\textbf {\bibinfo {volume}
  {289}},\ \bibinfo {pages} {137} (\bibinfo {year} {1992})}\BibitemShut
  {NoStop}%
\bibitem [{\citenamefont {Belitsky}\ \emph {et~al.}(2001)\citenamefont
  {Belitsky}, \citenamefont {Ji}, \citenamefont {Lu},\ and\ \citenamefont
  {Osborne}}]{trigluon_clar1}%
  \BibitemOpen
  \bibfield  {author} {\bibinfo {author} {\bibfnamefont {A.~V.}\ \bibnamefont
  {Belitsky}}, \bibinfo {author} {\bibfnamefont {X.}~\bibnamefont {Ji}},
  \bibinfo {author} {\bibfnamefont {W.}~\bibnamefont {Lu}},\ and\ \bibinfo
  {author} {\bibfnamefont {J.}~\bibnamefont {Osborne}},\ }\bibfield  {title}
  {\bibinfo {title} {{Singlet $g_2$ structure function in the next-to-leading
  order}},\ }\href {https://doi.org/10.1103/PhysRevD.63.094012} {\bibfield
  {journal} {\bibinfo  {journal} {Phys. Rev. D}\ }\textbf {\bibinfo {volume}
  {63}},\ \bibinfo {pages} {094012} (\bibinfo {year} {2001})}\BibitemShut
  {NoStop}%
\bibitem [{\citenamefont {Braun}\ \emph {et~al.}(2009)\citenamefont {Braun},
  \citenamefont {Manashov},\ and\ \citenamefont {Pirnay}}]{trigluon_clar2}%
  \BibitemOpen
  \bibfield  {author} {\bibinfo {author} {\bibfnamefont {V.~M.}\ \bibnamefont
  {Braun}}, \bibinfo {author} {\bibfnamefont {A.~N.}\ \bibnamefont
  {Manashov}},\ and\ \bibinfo {author} {\bibfnamefont {B.}~\bibnamefont
  {Pirnay}},\ }\bibfield  {title} {\bibinfo {title} {{Scale dependence of
  twist-three contributions to single spin asymmetries}},\ }\href
  {https://doi.org/10.1103/PhysRevD.80.114002} {\bibfield  {journal} {\bibinfo
  {journal} {Phys. Rev. D}\ }\textbf {\bibinfo {volume} {80}},\ \bibinfo
  {pages} {114002} (\bibinfo {year} {2009})}\BibitemShut {NoStop}%
\bibitem [{\citenamefont {Beppu}\ \emph {et~al.}(2010)\citenamefont {Beppu},
  \citenamefont {Koike}, \citenamefont {Tanaka},\ and\ \citenamefont
  {Yoshida}}]{trigluon_twists_Yoshida_SIDIS}%
  \BibitemOpen
  \bibfield  {author} {\bibinfo {author} {\bibfnamefont {H.}~\bibnamefont
  {Beppu}}, \bibinfo {author} {\bibfnamefont {Y.}~\bibnamefont {Koike}},
  \bibinfo {author} {\bibfnamefont {K.}~\bibnamefont {Tanaka}},\ and\ \bibinfo
  {author} {\bibfnamefont {S.}~\bibnamefont {Yoshida}},\ }\bibfield  {title}
  {\bibinfo {title} {{Contribution of twist-3 multigluon correlation functions
  to single spin asymmetry in semi-inclusive deep inelastic scattering}},\
  }\href {https://doi.org/10.1103/PhysRevD.82.054005} {\bibfield  {journal}
  {\bibinfo  {journal} {Phys. Rev. D}\ }\textbf {\bibinfo {volume} {82}},\
  \bibinfo {pages} {054005} (\bibinfo {year} {2010})}\BibitemShut {NoStop}%
\bibitem [{\citenamefont {Kang}\ and\ \citenamefont
  {Qiu}(2008)}]{trigluon_twists_SIDIS_kang}%
  \BibitemOpen
  \bibfield  {author} {\bibinfo {author} {\bibfnamefont {Z.-B.}\ \bibnamefont
  {Kang}}\ and\ \bibinfo {author} {\bibfnamefont {J.-W.}\ \bibnamefont {Qiu}},\
  }\bibfield  {title} {\bibinfo {title} {{Single transverse-spin asymmetry for
  $D$-meson production in semi-inclusive deep inelastic scattering}},\ }\href
  {https://doi.org/10.1103/PhysRevD.78.034005} {\bibfield  {journal} {\bibinfo
  {journal} {Phys. Rev. D}\ }\textbf {\bibinfo {volume} {78}},\ \bibinfo
  {pages} {034005} (\bibinfo {year} {2008})}\BibitemShut {NoStop}%
\bibitem [{\citenamefont {Adamczyk}\ \emph {et~al.}(2012)\citenamefont
  {Adamczyk} \emph {et~al.}}]{STAR_inclusivejet_TSSA1}%
  \BibitemOpen
  \bibfield  {author} {\bibinfo {author} {\bibfnamefont {L.}~\bibnamefont
  {Adamczyk}} \emph {et~al.} (\bibinfo {collaboration} {STAR Collaboration}),\
  }\bibfield  {title} {\bibinfo {title} {Longitudinal and transverse spin
  asymmetries for inclusive jet production at mid-rapidity in polarized
  $p\mathbf{+}p$ collisions at $\sqrt{s}=200$ gev},\ }\href
  {https://doi.org/10.1103/PhysRevD.86.032006} {\bibfield  {journal} {\bibinfo
  {journal} {Phys. Rev. D}\ }\textbf {\bibinfo {volume} {86}},\ \bibinfo
  {pages} {032006} (\bibinfo {year} {2012})}\BibitemShut {NoStop}%
\bibitem [{\citenamefont {Bland}\ \emph {et~al.}(2015)\citenamefont {Bland}
  \emph {et~al.}}]{AnDY_gluons}%
  \BibitemOpen
  \bibfield  {author} {\bibinfo {author} {\bibfnamefont {L.~C.}\ \bibnamefont
  {Bland}} \emph {et~al.} (\bibinfo {collaboration} {AnDY Collaboration}),\
  }\bibfield  {title} {\bibinfo {title} {{Cross Sections and Transverse
  Single-Spin Asymmetries in Forward Jet Production from Proton Collisions at
  $\sqrt{s}=500$ GeV}},\ }\href
  {https://doi.org/10.1016/j.physletb.2015.10.001} {\bibfield  {journal}
  {\bibinfo  {journal} {Phys. Lett. B}\ }\textbf {\bibinfo {volume} {750}},\
  \bibinfo {pages} {660} (\bibinfo {year} {2015})}\BibitemShut {NoStop}%
\bibitem [{\citenamefont {Adolph}\ \emph {et~al.}(2017)\citenamefont {Adolph}
  \emph {et~al.}}]{SIDIS_gluonSivers}%
  \BibitemOpen
  \bibfield  {author} {\bibinfo {author} {\bibfnamefont {C.}~\bibnamefont
  {Adolph}} \emph {et~al.} (\bibinfo {collaboration} {COMPASS Collaboration}),\
  }\bibfield  {title} {\bibinfo {title} {{First measurement of the Sivers
  asymmetry for gluons using SIDIS data}},\ }\href
  {https://doi.org/10.1016/j.physletb.2017.07.018} {\bibfield  {journal}
  {\bibinfo  {journal} {Phys. Lett. B}\ }\textbf {\bibinfo {volume} {772}},\
  \bibinfo {pages} {854} (\bibinfo {year} {2017})}\BibitemShut {NoStop}%
\bibitem [{\citenamefont {Adamczyk}\ \emph {et~al.}(2018)\citenamefont
  {Adamczyk} \emph {et~al.}}]{STAR_inclusivejet_TSSA2}%
  \BibitemOpen
  \bibfield  {author} {\bibinfo {author} {\bibfnamefont {L.}~\bibnamefont
  {Adamczyk}} \emph {et~al.} (\bibinfo {collaboration} {STAR Collaboration}),\
  }\bibfield  {title} {\bibinfo {title} {{Azimuthal transverse single-spin
  asymmetries of inclusive jets and charged pions within jets from
  polarized-proton collisions at $\sqrt{s} = 500$ GeV}},\ }\href
  {https://doi.org/10.1103/PhysRevD.97.032004} {\bibfield  {journal} {\bibinfo
  {journal} {Phys. Rev. D}\ }\textbf {\bibinfo {volume} {97}},\ \bibinfo
  {pages} {032004} (\bibinfo {year} {2018})}\BibitemShut {NoStop}%
\bibitem [{\citenamefont {Acharya}\ \emph
  {et~al.}(2021{\natexlab{a}})\citenamefont {Acharya} \emph {et~al.}}]{PPG234}%
  \BibitemOpen
  \bibfield  {author} {\bibinfo {author} {\bibfnamefont {U.}~\bibnamefont
  {Acharya}} \emph {et~al.} (\bibinfo {collaboration} {PHENIX Collaboration}),\
  }\bibfield  {title} {\bibinfo {title} {{Transverse single-spin asymmetries of
  midrapidity $\pi^0$ and $\eta$ mesons in polarized $p+p$ collisions at
  $\sqrt{s}=200$ GeV}},\ }\href {https://doi.org/10.1103/PhysRevD.103.052009}
  {\bibfield  {journal} {\bibinfo  {journal} {Phys. Rev. D}\ }\textbf {\bibinfo
  {volume} {103}},\ \bibinfo {pages} {052009} (\bibinfo {year}
  {2021}{\natexlab{a}})}\BibitemShut {NoStop}%
\bibitem [{\citenamefont {Abdallah}\ \emph {et~al.}(2022)\citenamefont
  {Abdallah} \emph {et~al.}}]{STAR_inclusive_jets}%
  \BibitemOpen
  \bibfield  {author} {\bibinfo {author} {\bibfnamefont {M.}~\bibnamefont
  {Abdallah}} \emph {et~al.} (\bibinfo {collaboration} {STAR}),\ }\bibfield
  {title} {\bibinfo {title} {{Azimuthal transverse single-spin asymmetries of
  inclusive jets and identified hadrons within jets from polarized $pp$
  collisions at $\sqrt{s}$ = 200 GeV}},\ }\href
  {https://doi.org/10.1103/PhysRevD.106.072010} {\bibfield  {journal} {\bibinfo
   {journal} {Phys. Rev. D}\ }\textbf {\bibinfo {volume} {106}},\ \bibinfo
  {pages} {072010} (\bibinfo {year} {2022})}\BibitemShut {NoStop}%
\bibitem [{\citenamefont {Aidala}\ \emph {et~al.}(2017)\citenamefont {Aidala}
  \emph {et~al.}}]{ohf_muons}%
  \BibitemOpen
  \bibfield  {author} {\bibinfo {author} {\bibfnamefont {C.}~\bibnamefont
  {Aidala}} \emph {et~al.} (\bibinfo {collaboration} {PHENIX Collaboration}),\
  }\bibfield  {title} {\bibinfo {title} {{Cross section and transverse
  single-spin asymmetry of muons from open heavy-flavor decays in polarized
  $p$+$p$ collisions at $\sqrt{s}=200$ GeV}},\ }\href
  {https://doi.org/10.1103/PhysRevD.95.112001} {\bibfield  {journal} {\bibinfo
  {journal} {Phys. Rev. D}\ }\textbf {\bibinfo {volume} {95}},\ \bibinfo
  {pages} {112001} (\bibinfo {year} {2017})}\BibitemShut {NoStop}%
\bibitem [{\citenamefont {Acharya}\ \emph
  {et~al.}(2021{\natexlab{b}})\citenamefont {Acharya} \emph {et~al.}}]{PPG235}%
  \BibitemOpen
  \bibfield  {author} {\bibinfo {author} {\bibfnamefont {U.}~\bibnamefont
  {Acharya}} \emph {et~al.} (\bibinfo {collaboration} {PHENIX Collaboration}),\
  }\bibfield  {title} {\bibinfo {title} {{Probing Gluon Spin-Momentum
  Correlations in Transversely Polarized Protons through Midrapidity Isolated
  Direct Photons in $p{\uparrow}$$+$$p$ Collisions at $\sqrt{s}=200$ GeV}},\
  }\href {https://doi.org/10.1103/PhysRevLett.127.162001} {\bibfield  {journal}
  {\bibinfo  {journal} {Phys. Rev. Lett.}\ }\textbf {\bibinfo {volume} {127}},\
  \bibinfo {pages} {162001} (\bibinfo {year} {2021}{\natexlab{b}})}\BibitemShut
  {NoStop}%
\bibitem [{\citenamefont {Kang}\ \emph {et~al.}(2008)\citenamefont {Kang},
  \citenamefont {Qiu}, \citenamefont {Vogelsang},\ and\ \citenamefont
  {Yuan}}]{trigluon_twists}%
  \BibitemOpen
  \bibfield  {author} {\bibinfo {author} {\bibfnamefont {Z.-B.}\ \bibnamefont
  {Kang}}, \bibinfo {author} {\bibfnamefont {J.-W.}\ \bibnamefont {Qiu}},
  \bibinfo {author} {\bibfnamefont {W.}~\bibnamefont {Vogelsang}},\ and\
  \bibinfo {author} {\bibfnamefont {F.}~\bibnamefont {Yuan}},\ }\bibfield
  {title} {\bibinfo {title} {{Accessing tri-gluon correlations in the nucleon
  via the single spin asymmetry in open charm production}},\ }\href
  {https://doi.org/10.1103/PhysRevD.78.114013} {\bibfield  {journal} {\bibinfo
  {journal} {Phys. Rev. D}\ }\textbf {\bibinfo {volume} {78}},\ \bibinfo
  {pages} {114013} (\bibinfo {year} {2008})}\BibitemShut {NoStop}%
\bibitem [{\citenamefont {Koike}\ and\ \citenamefont
  {Yoshida}(2011)}]{trigluon_twists_Yoshida}%
  \BibitemOpen
  \bibfield  {author} {\bibinfo {author} {\bibfnamefont {Y.}~\bibnamefont
  {Koike}}\ and\ \bibinfo {author} {\bibfnamefont {S.}~\bibnamefont
  {Yoshida}},\ }\bibfield  {title} {\bibinfo {title} {{Probing the three-gluon
  correlation functions by the single spin asymmetry in $p^\uparrow p \to
  DX$}},\ }\href {https://doi.org/10.1103/PhysRevD.84.014026} {\bibfield
  {journal} {\bibinfo  {journal} {Phys. Rev. D}\ }\textbf {\bibinfo {volume}
  {84}},\ \bibinfo {pages} {014026} (\bibinfo {year} {2011})}\BibitemShut
  {NoStop}%
\bibitem [{\citenamefont {Anselmino}\ \emph {et~al.}(2004)\citenamefont
  {Anselmino}, \citenamefont {Boglione}, \citenamefont {D'Alesio},
  \citenamefont {Leader},\ and\ \citenamefont {Murgia}}]{gluonSivers}%
  \BibitemOpen
  \bibfield  {author} {\bibinfo {author} {\bibfnamefont {M.}~\bibnamefont
  {Anselmino}}, \bibinfo {author} {\bibfnamefont {M.}~\bibnamefont {Boglione}},
  \bibinfo {author} {\bibfnamefont {U.}~\bibnamefont {D'Alesio}}, \bibinfo
  {author} {\bibfnamefont {E.}~\bibnamefont {Leader}},\ and\ \bibinfo {author}
  {\bibfnamefont {F.}~\bibnamefont {Murgia}},\ }\bibfield  {title} {\bibinfo
  {title} {{Accessing Sivers gluon distribution via transverse single-spin
  asymmetries in ${p}^{\ensuremath{\uparrow}}p\ensuremath{\rightarrow}DX$
  processes at BNL RHIC}},\ }\href {https://doi.org/10.1103/PhysRevD.70.074025}
  {\bibfield  {journal} {\bibinfo  {journal} {Phys. Rev. D}\ }\textbf {\bibinfo
  {volume} {70}},\ \bibinfo {pages} {074025} (\bibinfo {year}
  {2004})}\BibitemShut {NoStop}%
\bibitem [{\citenamefont {D'Alesio}\ \emph {et~al.}(2017)\citenamefont
  {D'Alesio}, \citenamefont {Murgia}, \citenamefont {Pisano},\ and\
  \citenamefont {Taels}}]{gluonSivers2}%
  \BibitemOpen
  \bibfield  {author} {\bibinfo {author} {\bibfnamefont {U.}~\bibnamefont
  {D'Alesio}}, \bibinfo {author} {\bibfnamefont {F.}~\bibnamefont {Murgia}},
  \bibinfo {author} {\bibfnamefont {C.}~\bibnamefont {Pisano}},\ and\ \bibinfo
  {author} {\bibfnamefont {P.}~\bibnamefont {Taels}},\ }\bibfield  {title}
  {\bibinfo {title} {{Probing the gluon Sivers function in $p^\uparrow p\to
  J/\psi\,X$ and $p^\uparrow p \to D\,X$}},\ }\href
  {https://doi.org/10.1103/PhysRevD.96.036011} {\bibfield  {journal} {\bibinfo
  {journal} {Phys. Rev. D}\ }\textbf {\bibinfo {volume} {96}},\ \bibinfo
  {pages} {036011} (\bibinfo {year} {2017})}\BibitemShut {NoStop}%
\bibitem [{\citenamefont {D'Alesio}\ \emph {et~al.}(2019)\citenamefont
  {D'Alesio}, \citenamefont {Flore}, \citenamefont {Murgia}, \citenamefont
  {Pisano},\ and\ \citenamefont {Taels}}]{gluonSivers3}%
  \BibitemOpen
  \bibfield  {author} {\bibinfo {author} {\bibfnamefont {U.}~\bibnamefont
  {D'Alesio}}, \bibinfo {author} {\bibfnamefont {C.}~\bibnamefont {Flore}},
  \bibinfo {author} {\bibfnamefont {F.}~\bibnamefont {Murgia}}, \bibinfo
  {author} {\bibfnamefont {C.}~\bibnamefont {Pisano}},\ and\ \bibinfo {author}
  {\bibfnamefont {P.}~\bibnamefont {Taels}},\ }\bibfield  {title} {\bibinfo
  {title} {{Unraveling the Gluon Sivers Function in Hadronic Collisions at
  RHIC}},\ }\href {https://doi.org/10.1103/PhysRevD.99.036013} {\bibfield
  {journal} {\bibinfo  {journal} {Phys. Rev. D}\ }\textbf {\bibinfo {volume}
  {99}},\ \bibinfo {pages} {036013} (\bibinfo {year} {2019})}\BibitemShut
  {NoStop}%
\bibitem [{\citenamefont {Schmidke}\ \emph {et~al.}(2018)\citenamefont
  {Schmidke} \emph {et~al.}}]{RHIC_polarimetry}%
  \BibitemOpen
  \bibfield  {author} {\bibinfo {author} {\bibfnamefont {W.~B.}\ \bibnamefont
  {Schmidke}} \emph {et~al.} (\bibinfo {collaboration} {The RHIC Polarimetry
  Group}),\ }\href {https://doi.org/10.2172/1473643} {\bibinfo {title} {{RHIC
  polarization for Runs 9--17}}} (\bibinfo {year} {2018}),\ \bibinfo {note}
  {{https://technotes.bnl.gov/Home/ViewTechNote/209057}}\BibitemShut {NoStop}%
\bibitem [{\citenamefont {Adcox}\ \emph
  {et~al.}(2003{\natexlab{a}})\citenamefont {Adcox} \emph
  {et~al.}}]{PHENIX_det}%
  \BibitemOpen
  \bibfield  {author} {\bibinfo {author} {\bibfnamefont {K.}~\bibnamefont
  {Adcox}} \emph {et~al.} (\bibinfo {collaboration} {PHENIX Collaboration}),\
  }\bibfield  {title} {\bibinfo {title} {{PHENIX detector overview}},\ }\href
  {https://doi.org/10.1016/S0168-9002(02)01950-2} {\bibfield  {journal}
  {\bibinfo  {journal} {Nucl. Instrum. Methods Phys. Res., Sec. A}\ }\textbf
  {\bibinfo {volume} {499}},\ \bibinfo {pages} {469} (\bibinfo {year}
  {2003}{\natexlab{a}})}\BibitemShut {NoStop}%
\bibitem [{\citenamefont {Nouicer}\ \emph {et~al.}(2008)\citenamefont {Nouicer}
  \emph {et~al.}}]{VTX_ref}%
  \BibitemOpen
  \bibfield  {author} {\bibinfo {author} {\bibfnamefont {R.}~\bibnamefont
  {Nouicer}} \emph {et~al.} (\bibinfo {collaboration} {PHENIX Collaboration}),\
  }\bibfield  {title} {\bibinfo {title} {{Silicon Vertex Tracker for PHENIX
  Upgrade at RICH: Capabilities and Detector Technology}},\ }\href
  {https://doi.org/10.22323/1.057.0042} {\bibfield  {journal} {\bibinfo
  {journal} {Proc. Sci.}\ }\textbf {\bibinfo {volume} {Vertex 2007}},\ \bibinfo
  {pages} {042} (\bibinfo {year} {2008})}\BibitemShut {NoStop}%
\bibitem [{\citenamefont {Nouicer}\ \emph {et~al.}(2009)\citenamefont {Nouicer}
  \emph {et~al.}}]{VTX_status}%
  \BibitemOpen
  \bibfield  {author} {\bibinfo {author} {\bibfnamefont {R.}~\bibnamefont
  {Nouicer}} \emph {et~al.} (\bibinfo {collaboration} {PHENIX Collaboration}),\
  }\bibfield  {title} {\bibinfo {title} {{Status and Performance of New Silicon
  Stripixel Detector for the PHENIX Experiment at RHIC: Beta Source,
  Cosmic-rays and Proton Beam at 120 GeV}},\ }\href
  {https://doi.org/10.1088/1748-0221/4/04/p04011} {\bibfield  {journal}
  {\bibinfo  {journal} {J. Instrum.}\ }\textbf {\bibinfo {volume} {4}}\bibinfo
  {number} { (04)},\ \bibinfo {pages} {P04011}}\BibitemShut {NoStop}%
\bibitem [{\citenamefont {Adcox}\ \emph
  {et~al.}(2003{\natexlab{b}})\citenamefont {Adcox} \emph
  {et~al.}}]{tracking_ref}%
  \BibitemOpen
\bibfield  {number} {  }\bibfield  {author} {\bibinfo {author} {\bibfnamefont
  {K.}~\bibnamefont {Adcox}} \emph {et~al.} (\bibinfo {collaboration} {PHENIX
  Collaboration}),\ }\bibfield  {title} {\bibinfo {title} {{PHENIX central arm
  tracking detectors}},\ }\href {https://doi.org/10.1016/S0168-9002(02)01952-6}
  {\bibfield  {journal} {\bibinfo  {journal} {Nucl. Instrum. Methods Phys.
  Res., Sec. A}\ }\textbf {\bibinfo {volume} {499}},\ \bibinfo {pages} {489}
  (\bibinfo {year} {2003}{\natexlab{b}})}\BibitemShut {NoStop}%
\bibitem [{\citenamefont {Aphecetche}\ \emph {et~al.}(2003)\citenamefont
  {Aphecetche} \emph {et~al.}}]{EMCal_ref}%
  \BibitemOpen
  \bibfield  {author} {\bibinfo {author} {\bibfnamefont {L.}~\bibnamefont
  {Aphecetche}} \emph {et~al.} (\bibinfo {collaboration} {PHENIX
  Collaboration}),\ }\bibfield  {title} {\bibinfo {title} {{PHENIX
  calorimeter}},\ }\href {https://doi.org/10.1016/S0168-9002(02)01954-X}
  {\bibfield  {journal} {\bibinfo  {journal} {Nucl. Instrum. Methods Phys.
  Res., Sec. A}\ }\textbf {\bibinfo {volume} {499}},\ \bibinfo {pages} {521}
  (\bibinfo {year} {2003})}\BibitemShut {NoStop}%
\bibitem [{\citenamefont {Aizawa}\ \emph {et~al.}(2003)\citenamefont {Aizawa}
  \emph {et~al.}}]{RICH_ref}%
  \BibitemOpen
  \bibfield  {author} {\bibinfo {author} {\bibfnamefont {M.}~\bibnamefont
  {Aizawa}} \emph {et~al.} (\bibinfo {collaboration} {PHENIX Collaboration}),\
  }\bibfield  {title} {\bibinfo {title} {{PHENIX central arm particle ID
  detectors}},\ }\href {https://doi.org/10.1016/S0168-9002(02)01953-8}
  {\bibfield  {journal} {\bibinfo  {journal} {Nucl. Instrum. Methods Phys.
  Res., Sec. A}\ }\textbf {\bibinfo {volume} {499}},\ \bibinfo {pages} {508}
  (\bibinfo {year} {2003})}\BibitemShut {NoStop}%
\bibitem [{\citenamefont {Aidala}\ \emph {et~al.}(2019)\citenamefont {Aidala}
  \emph {et~al.}}]{PPG223}%
  \BibitemOpen
  \bibfield  {author} {\bibinfo {author} {\bibfnamefont {C.}~\bibnamefont
  {Aidala}} \emph {et~al.} (\bibinfo {collaboration} {PHENIX Collaboration}),\
  }\bibfield  {title} {\bibinfo {title} {{Measurement of charm and bottom
  production from semileptonic hadron decays in $p+p$ collisions at
  $\sqrt{s_{NN}}=200$ GeV}},\ }\href
  {https://doi.org/10.1103/PhysRevD.99.092003} {\bibfield  {journal} {\bibinfo
  {journal} {Phys. Rev. D}\ }\textbf {\bibinfo {volume} {99}},\ \bibinfo
  {pages} {092003} (\bibinfo {year} {2019})}\BibitemShut {NoStop}%
\bibitem [{\citenamefont {Adare}\ \emph {et~al.}(2016)\citenamefont {Adare}
  \emph {et~al.}}]{PHENIX:2015ynp}%
  \BibitemOpen
  \bibfield  {author} {\bibinfo {author} {\bibfnamefont {A.}~\bibnamefont
  {Adare}} \emph {et~al.} (\bibinfo {collaboration} {PHENIX Collaboration}),\
  }\bibfield  {title} {\bibinfo {title} {{Single electron yields from
  semileptonic charm and bottom hadron decays in Au$+$Au collisions at
  $\sqrt{s_{NN}}=200$ GeV}},\ }\href
  {https://doi.org/10.1103/PhysRevC.93.034904} {\bibfield  {journal} {\bibinfo
  {journal} {Phys. Rev. C}\ }\textbf {\bibinfo {volume} {93}},\ \bibinfo
  {pages} {034904} (\bibinfo {year} {2016})}\BibitemShut {NoStop}%
\bibitem [{\citenamefont {Adare}\ \emph {et~al.}(2010)\citenamefont {Adare}
  \emph {et~al.}}]{PPG103E}%
  \BibitemOpen
  \bibfield  {author} {\bibinfo {author} {\bibfnamefont {A.}~\bibnamefont
  {Adare}} \emph {et~al.} (\bibinfo {collaboration} {PHENIX Collaboration}),\
  }\bibfield  {title} {\bibinfo {title} {{Measurement of Transverse Single-Spin
  Asymmetries for $J/\psi$ Production in Polarized $p+p$ Collisions at
  $\sqrt{s}=200$ GeV}},\ }\href {https://doi.org/10.1103/PhysRevD.82.112008}
  {\bibfield  {journal} {\bibinfo  {journal} {Phys. Rev. D}\ }\textbf {\bibinfo
  {volume} {82}},\ \bibinfo {pages} {112008} (\bibinfo {year} {2010})},\
  \bibinfo {note} {[Phys. Rev. D {\bf 86}, 099904(E) (2012)]}\BibitemShut
  {NoStop}%
\bibitem [{\citenamefont {Adler}\ \emph {et~al.}(2005)\citenamefont {Adler}
  \emph {et~al.}}]{PPG050}%
  \BibitemOpen
  \bibfield  {author} {\bibinfo {author} {\bibfnamefont {S.~S.}\ \bibnamefont
  {Adler}} \emph {et~al.} (\bibinfo {collaboration} {PHENIX Collaboration}),\
  }\bibfield  {title} {\bibinfo {title} {{Measurement of transverse single-spin
  asymmetries for mid-rapidity production of neutral pions and charged hadrons
  in polarized p+p collisions at $\sqrt{s}=200$ GeV}},\ }\href@noop {}
  {\bibfield  {journal} {\bibinfo  {journal} {Phys. Rev. Lett.}\ }\textbf
  {\bibinfo {volume} {95}},\ \bibinfo {pages} {202001} (\bibinfo {year}
  {2005})}\BibitemShut {NoStop}%
\bibitem [{\citenamefont {Sj\"ostrand}\ \emph {et~al.}(2001)\citenamefont
  {Sj\"ostrand}, \citenamefont {Eden}, \citenamefont {Friberg}, \citenamefont
  {L\"onnblad}, \citenamefont {Miu}, \citenamefont {Mrenna},\ and\
  \citenamefont {Norrbin}}]{Sjostrand:2000wi}%
  \BibitemOpen
  \bibfield  {author} {\bibinfo {author} {\bibfnamefont {T.}~\bibnamefont
  {Sj\"ostrand}}, \bibinfo {author} {\bibfnamefont {P.}~\bibnamefont {Eden}},
  \bibinfo {author} {\bibfnamefont {C.}~\bibnamefont {Friberg}}, \bibinfo
  {author} {\bibfnamefont {L.}~\bibnamefont {L\"onnblad}}, \bibinfo {author}
  {\bibfnamefont {G.}~\bibnamefont {Miu}}, \bibinfo {author} {\bibfnamefont
  {S.}~\bibnamefont {Mrenna}},\ and\ \bibinfo {author} {\bibfnamefont
  {E.}~\bibnamefont {Norrbin}},\ }\bibfield  {title} {\bibinfo {title}
  {{High-energy physics event generation with PYTHIA 6.1}},\ }\href
  {https://doi.org/10.1016/S0010-4655(00)00236-8} {\bibfield  {journal}
  {\bibinfo  {journal} {Comput. Phys. Commun.}\ }\textbf {\bibinfo {volume}
  {135}},\ \bibinfo {pages} {238} (\bibinfo {year} {2001})}\BibitemShut
  {NoStop}%
\end{thebibliography}

%
 
\end{document}